\documentclass{emulateapj}
\slugcomment{{\sc Accepted to AJ} October 4, 2018} 

\usepackage{color}
\usepackage{framed}
\usepackage{natbib}
\usepackage{textcomp}

\newcommand{\ms}{\ensuremath{\mathrm{m\,s}^{-1}}}

\newcommand{\mass}{\mathcal{M}}

\newcommand{\mearth}{\ensuremath{\rm M_{\oplus}}}
\newcommand{\rearth}{\ensuremath{\rm R_{\oplus}}}
\newcommand{\msini}{\ensuremath{\mass\sin i}}

\newenvironment{myitemize}
{ \begin{itemize} 
  \setlength{\itemsep}{0pt}
  \setlength{\parskip}{0pt}
  \setlength{\parsep}{0pt}   }
{ \end{itemize}         } 

\begin{document}

\title{Simulating the M-R Relation from APF follow up of {\it TESS} targets: \\Survey design and strategies for overcoming mass biases}

\author{Jennifer Burt\altaffilmark{1}, Brad Holden\altaffilmark{2}, Angie Wolfgang\altaffilmark{3,4}, L. G. Bouma\altaffilmark{5}}

\altaffiltext{1}{Kavli Institute for Astrophysics and Space Research, Massachusetts Institute of Technology, Cambridge, MA 02139, USA}
\altaffiltext{2}{UCO/Lick Observatory, University of California, Santa Cruz, 1156 High Street, Santa Cruz, CA 95064, USA}
\altaffiltext{3}{Center for Exoplanets and Habitable Worlds, 525 Davey Laboratory, The Pennsylvania State University, University Park, PA 16802, USA}
\altaffiltext{4}{Department of Astronomy and Astrophysics, The Pennsylvania State University, 525 Davey Laboratory, University Park, PA 16802, USA}
\altaffiltext{5}{Department of Astrophysical Sciences, Princeton University, 4 Ivy Lane, Princeton, NJ 08540, USA}

\begin{abstract}

We present simulations of multi-year RV follow up campaigns of the {\it TESS} small exoplanet yield on the Automated Planet Finder telescope, using four different schemes to sample the transiting planets' RV phase curves. For planets below roughly 10 M$_\oplus$ we see a systematic bias of measured masses that are higher than the true planet mass, regardless of the observing scheme used. This produces a statistically significant difference in the mass-radius relation we recover, where planet masses are predicted to be too high and too similar across the entire super-Earth to Neptune radius range. This bias is due in part to only reporting masses that are measured with high statistical significance. Incorporating all mass measurements, even those that are essentially only upper limits, significantly mitigates this bias. We also find statistically significant differences between the mean number of planets measured at the 1-, 3-, and 5$\sigma_{\rm K}$ level by the different prioritization schemes. Our results show that prioritization schemes that more evenly sample the RV phase curves produce a larger number of significant mass detections. The scheme that aims to most uniformly sample the phase curve performs best, followed closely by the scheme which randomly samples, and then an in quadrature sampling approach. The fourth scheme, out of quadrature, performs noticeably worse. These results have important implications for determining accurate planet compositions and for designing effective RV follow up campaigns in the era of large planet detection surveys such as $K2$, $TESS$, and $PLATO$.


\end{abstract}

\section{Introduction}

The super-Earth and sub-Neptune planets detected by NASA's {\it Kepler} mission have asserted their position as the most common planet archetype in the galaxy, at least for planets with periods less than 100 days. These planets, with radii ranging from 1-4 \rearth, make up 75\% of the final catalog of {\it Kepler} exoplanet candidates (GAIA DR25; \citealt{Thompson2018}). Despite their ubiquity, however, we do not yet have a clear picture of the mass-radius relation that governs these objects. They span the transition in radius space from small, predominantly rocky bodies to planets with the voluminous layers of volatiles seen in our own ice giants. Thus even planets with the same radius can have masses that vary widely due to the large range of possible compositions \citep{Wolfgang2015}, e.g. rock, astrophysical ices, and H/He gases. 

{\it Kepler} has detected thousands of exoplanet candidates with radii ranging from 1-4 \rearth\ \citep{Akeson2013,Coughlin2016,Thompson2018}, yet providing well constrained masses for them has proven non-trivial. Many of these planets orbit faint stars and impart radial velocity (RV) signals of \textless\ 1\ms. As only the most advanced RV telescopes currently in operation have consistently demonstrated 1 \ms\ precision on bright, quiet, stars, follow up efforts to determine the {\it Kepler} planets' masses are both challenging and time intensive. Indeed, as of September 2018, only 90 of the 2327 confirmed, transiting {\it Kepler} planets have had both their radii and masses determined beyond upper limits via RV measurements. The second phase of \textit{Kepler}, {\it K2}, is providing additional RV-amenable targets as it searches the ecliptic plane for evidence of transits around brighter stars in 80 day observing campaigns. As of September 2018, {\it K2} has released data from 18 different campaigns resulting in 325 confirmed, transiting planets, 37 of which have had both their radii and masses determined beyond upper limits via RV measurements\footnote{ Based on data from the NASA Exoplanet Archive, accessed on September 24th, 2018 \url{https://exoplanetarchive.ipac.caltech.edu}.}.

\begin{figure}
\centering
\plotone{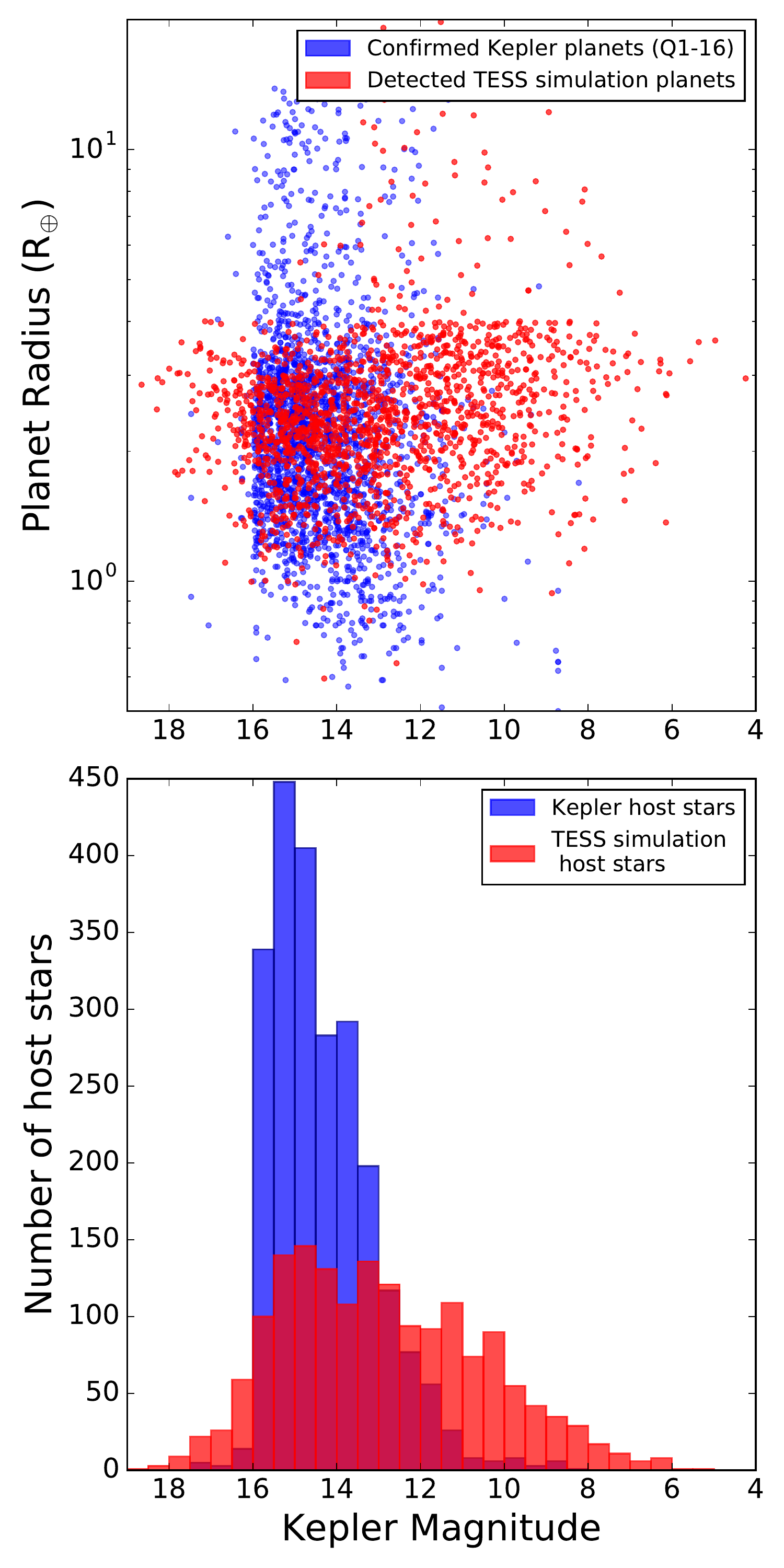}
\caption{\label{fig:Kepler_TESS} {\it Top} : Comparisons of the confirmed {\it Kepler} planets from Q1-Q16 \citep{Mullally2015} in blue and the simulated planets from our instance of the predicted {\it TESS} planet yield in red. {\it Bottom}: {\it Kepler} magnitudes of the detected {\it Kepler} host stars (blue) and simulated detected {\it TESS} planet host stars (red).}
\end{figure}

While RV instrumental precision and capabilities are improving \citep{Fischer2016, WrightRobertson2017}, perhaps the best hope for further populating the mass-radius diagram with small exoplanets in the near term lies with NASA's {\it TESS} mission (the Transiting Exoplanet Survey Satellite), which started returning science data in August 2018 \citep{Ricker2014}. {\it TESS} will survey the entire night sky, looking for evidence of exoplanet transits across the closest, brightest stars. The mission's observing strategy, however, means that $\sim$65\% of the sky will be observed for only 27 days during the primary mission, which will not provide a long enough baseline of {\it TESS} photometry to search for evidence of transit timing variations. Early analysis suggests that {\it TESS} will likely find only $\sim$10 planets with strong TTVs \citep{Kane2017}. Fortunately, many of the stars observed using the satellite's two minute cadence mode will be brighter than V = 12, making them plausible targets for ground based RV mass measurements \citep[hereafter S15, see also Figure \ref{fig:Kepler_TESS}]{Sullivan2015}. 

{\it TESS} will provide a large step forward in our efforts to understand the M-R relation in small exoplanets, providing 100+ bright stars (V \textless\ 12) with transiting planet candidates, and many more candidates around faint stars (12 \textless\ V \textless\ 16) (S15). Indeed only a month after the sector one {\it TESS} exoplanet candidate alerts were released, two {\it TESS} planets have already been confirmed. The first {\it TESS} planet, Pi Mensae c, is a 2 \rearth\ planet in a six day orbit that imparts a $K$ = 1.6 \ms\ reflex motion on its $V$=5.6 host star \citep{Huang2018, Gandolfi2018}. The second {\it TESS} planet, LHS 3844 b, is a 1.4 \rearth\, ultra-short period planet orbiting a nearby M dwarf \citep{Vanderspek2018}. 

These early confirmations are exactly the kinds of planets that {\it TESS} was designed to find - small, short period planets around nearby stars - and suggest that many more exciting planets are soon to come. With such a plethora of targets, RV follow up efforts across different telescopes, and even across different nights/seasons but on the same telescope, need to be well thought out and coordinated. This requires developing a detailed understanding of the potential contributions of each RV facility to the M-R diagram, specifically which areas of parameter space they can fill in and what their target coverage and error bars are likely to look like after a given amount of observing time. 

With the transit emphemerides that missions such as {\it TESS} and {\it Kepler} provide, RV follow up can be optimized to sample the expected RV phase curves in a variety of different ways. The most rapid and low-cost approach in terms of telescope time would be to take a smaller number of observations at or near the planet's quadrature phases when the star's reflex motion is maximized. This method has, however, been shown to have certain biases especially in cases where the orbit is non-circular \citep{Ford2008, Loredo2011}. An approach that attempts to sample each RV phase curve as uniformly as possible seems promising if transit ephemerides are readily available, but also raises concerns of oversampling the curve and being inefficient. Performing an investigation into how different kinds of observing schemes compare, both in terms of the masses they measure and their resulting mass-radius relation implications, is key to making sure that the limited number of precise RV follow instruments currently in operation are used efficiently and effectively in the era of {\it TESS}.

In this paper, we investigate the effects of four different follow up observation schemes on the recovered planet masses and compositions of simulated {\it TESS} as measured by the Automated Planet Finder (APF) telescope. We begin in \S \ref{sect:methodology} by describing the creation of our simulated planet candidate database, and our use of a probabilistic mass-radius relation to account for composition-induced variation when identifying likely masses given simulated radii and when setting the targets' desired RV precisions. \S \ref{sect:newpriorities} details the design and implementation of four new, time-varying, prioritization schemes that interface with the APF's dynamic scheduler to plan the telescope's minute-to-minute observing strategy. Then in \S \ref{sect:simulator} we describe the observing simulator we have created for the APF and the details of our simulated, multi-year, radial velocity follow up campaigns. We analyze the resulting semi-amplitude and mass measurements in \S \ref{sect:results}, before examining how our observations will impact the M-R diagram and discussing the biases that we recover in the underlying M-R relation in \S \ref{sect:MR_Results}. Finally, we conclude in \S \ref{sect:conclusion} with a discussion of how the M-R diagram is likely to evolve with the advent of {\it TESS}, and how the mass bias we uncover in our own results could already be impacting planet confirmation and characterization efforts.
 
\section{Methodology} \label{sect:methodology}

\subsection{The Automated Planet Finder} \label{sect:APF}
The APF is a 2.4m telescope located at the Mt. Hamilton station of UCO/Lick observatory. The telescope is coupled with the high resolution (R$_{\rm max} \sim$ 120,000), prism cross-dispersed Levy echelle spectrograph, which covers a wavelength range of $\sim$3700 - 9700$\AA$. A full description of the design and the individual components of the APF is available in \citet{Vogt2014a}. We measure the Doppler shift of a star by imprinting an iodine absorption spectrum on the incident starlight before it passes through the spectrograph slit \citep{Butler1996}. The forest of added I$_{\rm 2}$ lines generates a stable wavelength calibration from 5000 to 6200 $\AA$ (the ``iodine region'') and permits the measurement of the spectrometer point spread function (PSF). For each spectrum so obtained, the iodine region is then divided into 700 individual, 2$\AA $ chunks, with each chunk providing an independent measure of the wavelength, the PSF, and the Doppler shift. The reported radial velocity from a given stellar spectrum is a the mean of all 700 individual RV measurements. The uncertainty for each velocity is the RMS of the individual segments' velocity values about the mean, divided by the square root of the number of segments. This ``internal'' uncertainty represents primarily errors in the fitting process, which are dominated by Poisson statistics. 

To support long-running RV surveys, we designed and implemented a dynamic scheduler capable of running the telescope without human interaction. The scheduler combines current observing data (atmospheric seeing, cloud cover), target star parameters (V magnitude, B-V color, expected jitter, current elevation) and survey design parameters (desired RV precision, desired cadence) to provide real time decisions on the optimal target for the telescope to observe. This automation and optimization increases observing efficiency, decreases operating costs and minimizes the potential for human error \citep[][hereafter B15]{Burt2015}. During the development of this scheduler, we also constructed a full telescope simulator suite to allow for testing new scheduler modifications. This software provides the scaffolding for the numerous simulations described throughout the current work.

\subsection{{\it TESS} Simulation \& APF target selection} \label{sect:targetselection}

S15 constructed a model of the local stellar and planetary populations, and combined it with a model for {\it TESS}'s photometric precision to predict the properties of the planetary systems that {\it TESS} is likely to detect. For the stars, they used the TRILEGAL stellar population synthesis model \citep{Girardi2005}, but replaced the TRILEGAL absolute magnitudes, T$_{\rm effs}$, and radii with those of the more recent Dartmouth models \citep{Dotter2008}. This was 
necessary in order to bring the radius-magnitude relation into agreement with observed interferometric radii \citep{Boyajian2012a,Boyajian2012b}. For the planets, they used occurrence rates derived from the {\it Kepler} mission: for host stars with $T_{\rm eff}>4000\,{\rm K}$ they used the rates calculated by \citet{Fressin2013}. Below this threshold, they used those found by \citet{DressingCharbonneau2015}. 

In this work, we employ a predicted {\it TESS} planet detection catalog similar to S15's Table 6, produced by a near-identical code to that which was used by S15. The main difference is that for any planetary system with at least one {\it TESS}-detected planet, our catalog includes all other simulated planets in that system, regardless of their detectability. This catalog has about 1670 detected planets with radii $R_p<4\rearth$ (Figure \ref{fig:Kepler_TESS}), which represent the detections expected from the $\sim 2\times10^5$ target stars observed at two minute cadence, selected for small-planet detectability (these two minute targets are also known as ``postage stamp stars''). This planet catalog does not include simulated detections from the {\it TESS} Full Frame Images (FFIs), which are created by stacking the exposures taken across all 4 cameras every thirty minutes. Imposing the same cuts we make to the postage stamp catalog, we find that the FFI detections increase the total number of planets we could follow up by $\sim$10\%, but that most of these targets are around stars on the faint end of the APF's performance range. Thus we choose to base our survey only on the postage stamp stars, although the FFI planets could be surveyed at higher cadence during potential {\it TESS} extended missions \citep{Bouma2017}.

As in the original S15 simulations, multiple planet systems are assigned to stars via independent draws from consecutive period-radius occurrence bins. The only requirements are that: 1) the periods of adjacent planetary orbits must have ratios of at least 1.2 as is generally seen in the {\it Kepler} data \citep[see e.g. Figure 4 in][]{Fabrycky2014}, and 2) planets around a star with a binary companion cannot have orbital periods that are within a factor of 5 of the binary orbital period \footnote{This ratio was chosen to be a bit more conservative that what can be derived from Equation 1 of \citet{Holman1999}}. The result is that in this instance of the simulation, 53\% of the transiting systems around FGK stars and 55\% of those around M stars are multiple-planet systems.

S15's simulation also assumed that all planets are on circular orbits around their host stars. We keep this convention throughout our simulation and analysis efforts, as modifying the eccentricity values could alter the {\it TESS} detection results. One of the aims of this work, however, is to compare optimal strategies for measuring planet masses, and the effectiveness of our various observing approaches would likely change with planetary eccentricity. We discuss this issue further in Sec.~\ref{sect:massmeasurements}. We additionally assume that all orbits are both coplanar and prograde because of the observed low mutual inclinations of transiting multiple planet systems \citep[e.g.][]{TremaineDong2012, Figueira2012, Fabrycky2014} and the fact that retrograde orbits have only been observed for Hot Jupiters, which make up a small component of the exoplanet population \citep{Wang2015}.

Another systematic limitation is that S15's procedure for constructing multiple planet systems ignored the correlations that exist between {\it Kepler}'s occurrence and multiplicity distributions. For instance, in a {\it Kepler} multi, in any given planet pair the outer planet is likely to be larger \citep[e.g.][]{Ciardi2015, Weiss2017}. Similarly, if a system has a short-period transiting planet, it is much more likely to host outer transiting planets \citep[e.g.][]{KippingLam2017}. These unaddressed modeling problems limit our ability to predict the number of systems with planets missed by {\it TESS} but detectable through RV follow up, and thus in this work we focus only on measuring masses for the planets actually detected by {\it TESS}.

{\it TESS}'s short observing lifetime, combined with the fact that the transit probability of a planet scales inversely with its semi-major axis, makes it unlikely that the mission will observe the multiple transit events required to flag any long period, Jovian planets that orbit {\it TESS} target stars. Such planets will almost certainly be detected via other analysis methods, but even then we will not be able to measure their orbital periods. These planets, however, will still impart notable signals in RV surveys thanks to the large semi-amplitudes they induce on their host stars. Our own Jupiter, for example, produces a $\sim$10\ms\ signal on the sun. To address the existence of such planets, we adopt the true Jupiter analog occurrence rate from \citet{Rowan2016}, and insert Jupiter analogs into 3\% of the stars in the {\it TESS} simulation catalog. Following the definitions presented in \citet{Rowan2016}, these analogs have periods between 5 and 15 years, and have masses between 0.3 and 3 M$_{\rm Jup}$. We draw the characteristics of our Jovian planets from a uniform distribution between the end points of each parameter because the population of known Jupiter analogs is still quite small (there are 8 in total, as of \citet{Rowan2016}) which means that determining detailed distribution shapes for the period, mass, etc is still beyond our reach, and pick the systems that these giant planets are inserted into at random. Following the procedure of S15, we put these planetary additions on circular and prograde orbits around their host stars in the same plane as the transiting planets detected by {\it TESS}.

From the resulting exoplanet host star catalog, we cull the stars that are:

\begin{myitemize}
\item physically visible to the APF (dec\textgreater -15)
\item bright enough for the APF to perform well (V\textless 11)
\item likely F/G/K/M dwarfs (T$_{\rm eff}$ \textless\ 6000K)
\end{myitemize}

For stars that satisfy these criteria, we assign rotation periods using the results of \citet{McQuillan2014}, which measured the rotation periods of 34,000 Kepler main sequence stars with T$_{\rm eff}$ \textless\ 6500 K. For each star in our simulation, we draw a rotation period at random from the subset of the McQuillan targets that have effective temperatures within 100 K of the TESS star. This period is used both for designing one of the stellar noise components (see Section \ref{sect:jitter}) and for discarding stars that are fast rotators and thus poor candidates for RV analysis due to rotational line broadening.

The traditional metric for deciding whether a star is a fast rotator is v$\sin{i}$, but here instead we calculate a basic rotational velocity by combining the stellar rotational periods with the stars' radii, as provided in the input {\it TESS} catalog. Stars with v$_{\rm rot}$ \textgreater\ 5 km/s are excluded from the survey target list, replicating the criteria used in the actual Lick-Carnegie exoplanet survey that produced the APF data set our scheduler is based on. This serves as a conservative estimate as the inclusion of the $\sin{i}$ term can only narrow the spectral features, thereby increasing the RV information content \citep{Bouchy2001, Beatty2015}.

\subsection{Assigning stellar noise}
\label{sect:jitter}

Stellar jitter is one of the key sources of noise in radial velocity data, and understanding its effects is important for long term RV surveys \citep{Lovis2011}. We use the work of \citet{Isaacson2010} (hereafter IF10) to generate a jitter value for each of the simulated {\it TESS} stars in our survey. IF10 finds that stellar jitter depends strongly on a star's B-V color, and gives a relation for four different B-V color bins (0.7 \textless B-V \textless 0.7, 0.7 \textless B-V \textless 1.0, 1.0 \textless B-V \textless 1.3, and 1.3 \textless B-V \textless 1.6). Within each bin, the authors fit the jitter as a function of activity as measured by the star's S index \citep{Duncan1991}. This activity metric is a differential measurement, $\Delta$S, measured from the baseline level for quietest stars in each color bin. For our catalog, we pick a median stellar activity level for each star based on its corresponding IF10 bin.

Once a given star has been assigned a total jitter value (jit$_{\rm tot}$), we divide that value into three contributions that sum to jit$_{\rm tot}$ when added in quadrature. One component is a random value modeled as a normal distribution, used to simulate the true, white noise ``jitter'' part of stellar activity. The other two components are sinusoidal, periodic signals, and are used to roughly simulate the p-mode oscillations and stellar rotation. The p-mode period is between 5 and 15 minutes \citep[based on measurements of the Sun:][]{SchrijverZwann2000, Broomhall2009}, while the star's rotational period is based on analogous {\it Kepler} stars as described in Section \ref{sect:targetselection}. Each of these periodic components is also assigned an initial, random phase at the beginning of the simulation. For each simulated observation,  we calculate the phase of the observation with respect to the starting phase and period so the signals will be periodic. The amplitude, however, is a random positive deviate. Depending on the phase of observation with respect to that of the two periodic noise sources, the actual deviate for an observation can be either positive or negative.

\subsection{Calculating planetary masses}\label{sect:planetmasses}

The updated S15 simulations provide planet radii for the generated {\it TESS} population. To simulate RV follow up of these planets, we need a way to map these radii to planet masses. Specifically, we require an M-R relation that can account for both the observed astrophysical variation in planet masses at a given radius and the uncertainty in this relation due to our data-starved, pre-{\it TESS} understanding of the underlying M-R distribution. The relation derived by \citet{Wolfgang2016} (hereafter WRF16) allows us to do both: its probabilistic nature accounts for the first concern, and its Bayesian formulation facilitates the second. In particular, we use the version of the M-R relation fit to planets between 1 and 8\rearth\ (see Eqn \ref{eqn:MR}, where $\sim$ means ``drawn from the distribution''). The parameters of this relation, taken from the 6th line of WRF16's Table 1, are the normalization constant, C = 1.6, the power-law index, $\gamma$ = 1.8, and the intrinsic scatter, $\sigma_{\rm M}$ = 2.9. We then extrapolate this relation up to 9\rearth, and finally use a lognormal distribution with $\mu(\textrm{log}(M/M_{\rm Jup}))=0.046$ and $\sigma(\textrm{log}(M/M_{\rm Jup}))=0.392$ to determine the masses of planets with radii, R \textgreater\ 9\rearth because the sample creates a skewed distribution that looks normal in log space. This last distribution was fit via a standard maximum likelihood method to the mass distribution of all Jupiters that appeared in the Exoplanet Orbit Database \citep{Han2014} as of December 3, 2015; it therefore assumes the observed, period-marginalized, Jupiter mass distribution is complete and unbiased.

\begin{equation}\label{eqn:MR}
\frac{M}{M_\oplus} \sim\ \rm Normal\left(\mu = C \left(\frac{R}{R_\oplus}\right)^\gamma ,\sigma = \sigma_M \right)
\end{equation}

\begin{figure}
\centering
\plotone{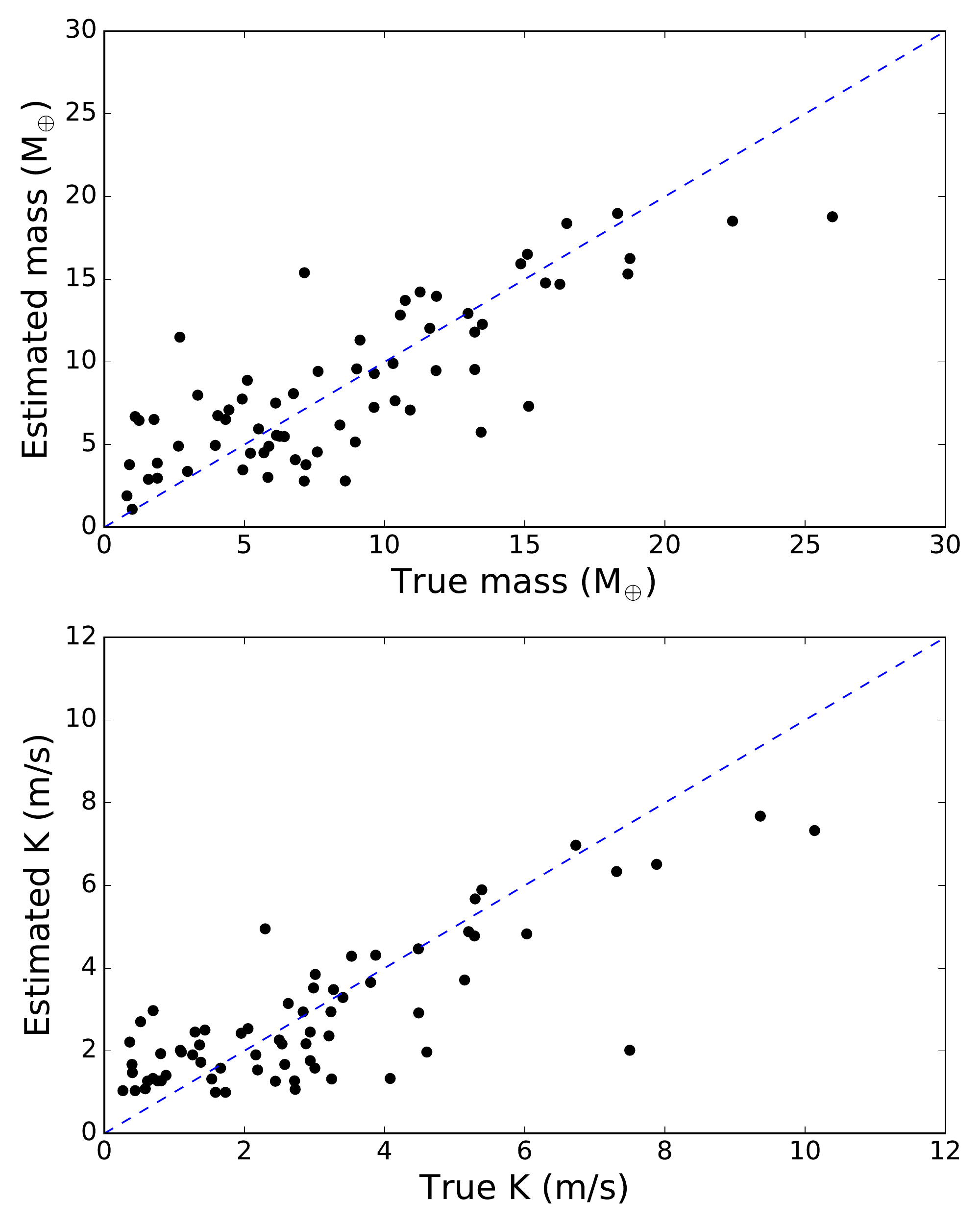}
\caption{\label{fig:MassHist} {\it Top}: The true and estimated mass values for each of the {\it TESS} simulation planets orbiting the culled {\it TESS} simulation stars. This excludes the true Jupiter analogs we add to the simulation, as we only have true masses for them, and no estimated masses. {\it Bottom}: Same as above, but showing the true and estimated semi-amplitude values for each of the {\it TESS} simulation planets orbiting the culled {\it TESS} simulation stars.}
\end{figure}

This M-R relation serves two purposes in our simulations. First, we assign ``true'' masses ($M_{\rm true}$) to the planets in the generated {\it TESS} population, which are used to create our simulated RV data. In particular, we marginalize over the posteriors of the M-R relation parameters while generating each planet's $M_{\rm true}$ in order to include the data-driven uncertainty in the relation (see the discussion about the posterior predictive distribution in Section 6.1 of WRF16 for more details).

However, having $M_{\rm true}$ in-hand for these planets does not accurately represent the information we will have as we initiate RV follow up of {\it TESS} planets. At that point we will not know $M_{\rm true}$ --- indeed, measuring it is the entire point of RV follow up efforts. Rather, we will assign each planet an estimated mass based upon its {\it TESS} derived radius of $M_{\rm est} = 1.6 R^{1.8}$ (the ``mean'' relation reported in WRF16) in order to decide what RV precision the APF should aim to achieve on each star. $M_{\rm est}$ in turn determines the observing feasibility of specific {\it TESS} planets each night based on the exposure times needed to achieve the required precision (see Section 4.1). Since consistently high-priority planets will end up populating the observed M-R distribution, and therefore influence the inferences we make about the true underlying mass distribution, $M_{\rm est}$ plays an important role in our simulations. 

The average difference between $M_{\rm true}$ and $M_{\rm est}$ for the planets in our sample is 2.3M$_\oplus$ (see Figure \ref{fig:MassHist} for the resulting mass distributions.). In general for normal distributions, the mean absolute deviation is $\sqrt{2/\pi}$ times the standard deviation \citep{Geary1935}, and so this result is consistent with the standard deviation of the input M-R relation (2.9M$_\oplus$). 

\subsection{Setting desired RV precisions}\label{sect:setprec}

After determining $M_{\rm true}$ and $M_{\rm est}$, we calculate the estimated RV semi-amplitude each planet imparts on its host star using Kepler's 3rd law (Equation 2) and taking $M_{\rm est}$ as the mass of the planet.

\begin{equation}
K_{{\rm est}} = \left(\frac{2 \pi G}{P}\right)^{1/3} \frac{M_{\rm est} \ \sin{i}}{\left(M_{\rm star}+M_{\rm est}\right)^{2/3}}
\end{equation}\label{eqn:RVamp}

\begin{figure}
\centering
\plotone{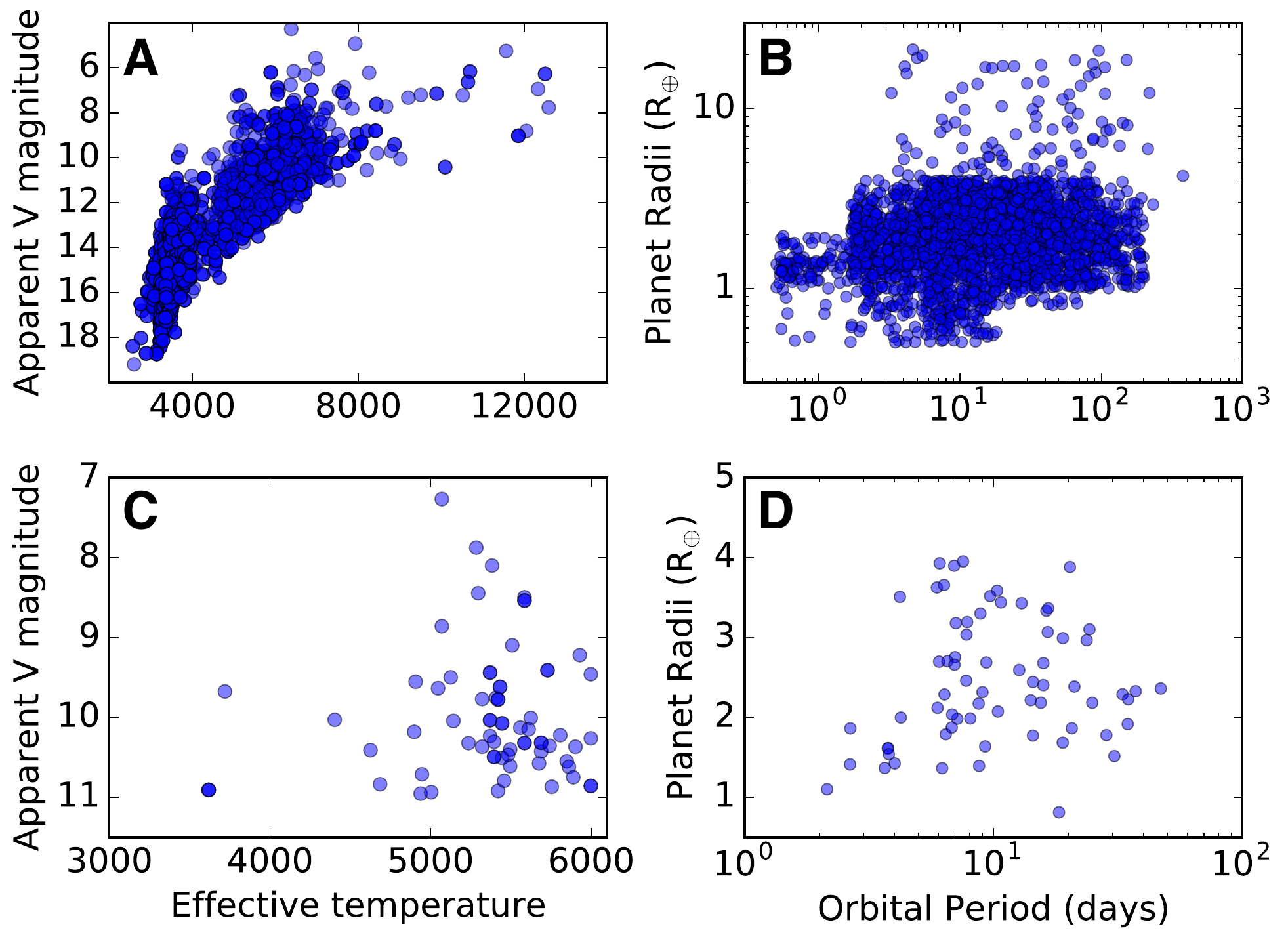}
\caption{\label{fig:TESS_selections} {\it A\ }: Effective temperature (T$_{\rm eff}$) vs. apparent V magnitude for the stars in the {\it TESS} simulation. {\it B\ }: Orbital period vs. planetary radius for all planets in the {\it TESS} simulation, whether or not they were detected by {\it TESS}. {\it C\ }: T$_{\rm eff}$ vs. apparent V magnitude for the stars included in our APR RV follow up program (V \textless 11, T$_{\rm eff}$ \textless 6000K, and Dec \textgreater -15). {\it D\ }: Orbital period vs. planetary radius for planets orbiting stars meeting our selection criteria for the APF, whether or not they were detected by {\it TESS}, plus the true Jupiter analogs we add around 3\% of the simulation stars.}
\end{figure}

Stars where no planet is expected to produce an RV semi-amplitude of K $\geq$ 1\ms\ are removed from the survey, in an effort to spend our observing time on systems where the APF is most likely to provide well measured masses. This is the last criteria imposed on our input target list, and the surviving host stars and planet candidates are shown in Figure \ref{fig:TESS_selections}. The target list includes 71 planets around 58 different stars.

\begin{figure}
\centering
\plotone{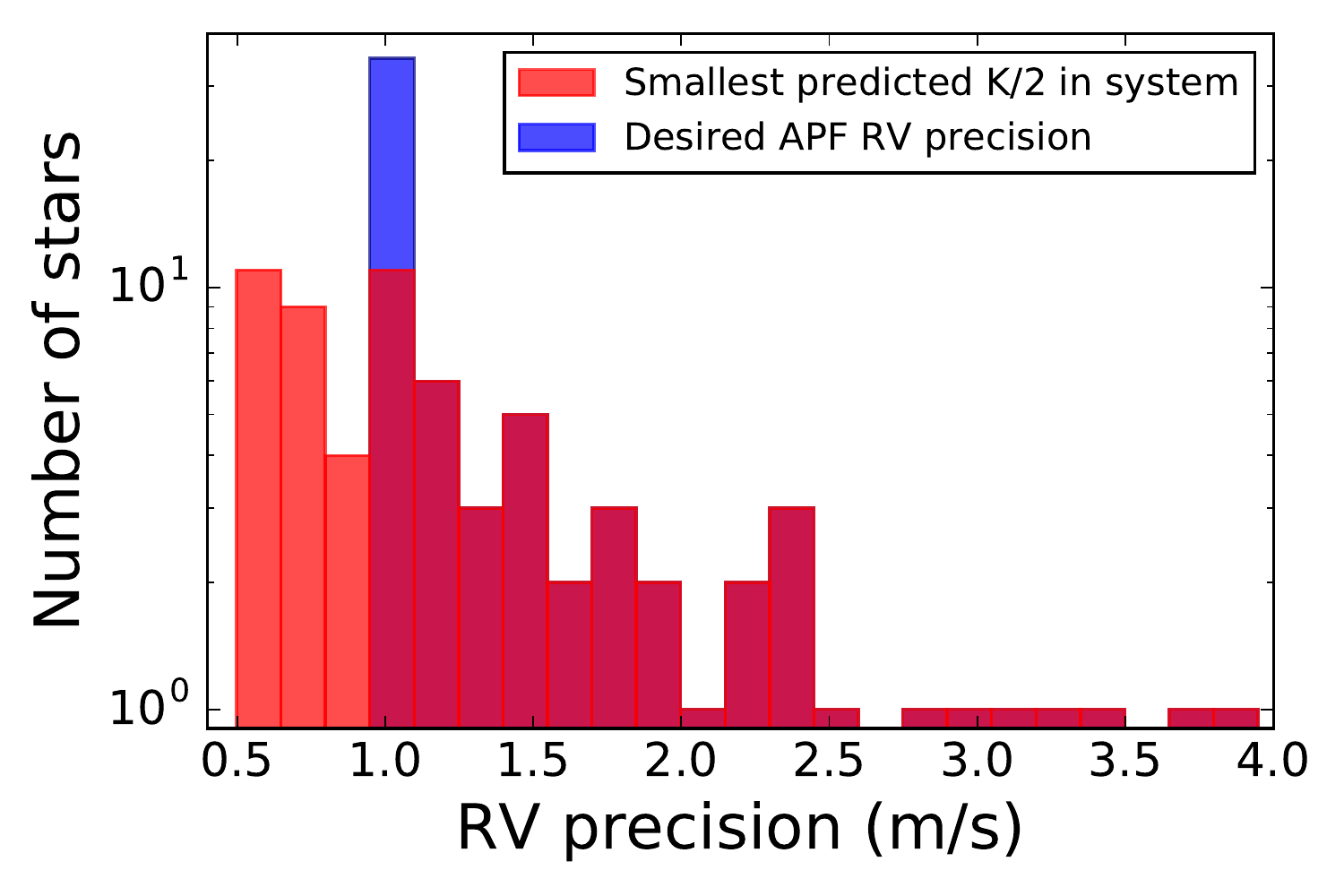}
\caption{\label{fig:PrecHist} Histogram of the K/2 values (in light red) and resulting desired precision values set for each of the culled {\it TESS} simulation stars (in blue). The overlap region appears as a dark red. These values are obtained by calculating the $K$ values of any planets detected by {\it TESS} in the simulation where the smallest semi-amplitude is at least 1\ms\ and then taking the smallest semi-amplitude in each system and dividing it in half.  For those stars whose result is \textless\ 1\ms, we instead set the desired precision level to 1\ms so as not to come up against the noise floor of the APF.}
\end{figure}

We then calculate an idealized RV precision level for all stars in our survey, which is set to be half of the planet's estimated semi-amplitude (K/2). This value comes from a desire to efficiently determine whether or not a given planet's Keplerian signal is actually visible in the RV data, which may or may not be the case based on how close M$_{\rm est}$ is to M$_{\rm true}$. In the ideal circumstance of no other noise sources, a precision of K/2 on each RV measurement would produce a detection at the 6-$\sigma_{\rm K}$ level within 10 exposures - a data set which is easily obtainable during one semester. In actuality, the variety of other noise sources present in real world RV data and simulated here will suppress the final signal to noise value for any target, and we will need more than 10 exposures to determine if the planet is visible in the data. For each star, we then compare its ideal RV precision level to the APF's precision floor of $\sim$1.0\ms (B15), and set the scheduler's ``desired precision'' criteria equal to the larger of the two values (Figure \ref{fig:PrecHist}).

\section{Creating time-dependent observing priorities} \label{sect:newpriorities}

One common approach to traditional RV surveys \citep[e.g.][]{Teske2016, Butler2017}, is to assign every star an observing priority that signifies how interesting that star is to the science team, and which acts as a weighting factor when deciding what stars to observe during the night. These priorities are then reassessed on a timescale of weeks/months/years to reflect changes in the target's interest level due to emerging, potentially Keplerian, signals. Between review sessions, however, these static observing priorities weight the star with the same level of observing desirableness at all times.

This is a sensible (and successful, see e.g. \citep{Vogt2015, Fulton2015, Bonfils2013} approach when running an RV survey without prior knowledge of the planets orbiting a star, as any observation could provide crucial insights to the star's (potentially planet-induced) movements. When performing follow up observations for transit missions like {\it TESS}, however, the photometry data set provides important {\it a priori} information that allows the development of more finely tuned, time varying priority values for target stars. The key to this approach is the fact that the RV phase curve of a transiting planet can be mapped out using the planet's orbital phase (calculated from its period and epoch of midtransit as provided by {\it TESS}) and instantaneous radial velocity (inferred from the planet's radius as provided by {\it TESS} and the corresponding mass value we assign in Section \ref{sect:planetmasses}). This assumes that the transit ephemeris for the planet originally established by {\it TESS} has not gone stale, despite the fact that they can decay quickly over the course of weeks to months \citep{Benneke2017, Bouma2017}. But for the small planets orbiting bright stars that are the focus of TESS's level one science requirement, it is likely safe to assume that they will undergo extensive photometric follow up which will help to stabilize the transit ephemerides over the few year time scale modeled here. Thus, instead of observing blindly throughout a planet's orbit, we can use the {\it TESS} light curves and any additional photometric follow up data to make informed decisions on which locations in the planet's RV phase curve to observe in order to optimize the impact of a given exposure. 

The instantaneous radial velocity for a given planet at time $t$ is given by Equation 3, where $f$ is the true anomaly, $\omega$ is the argument of periastron, $K$ is the RV semi-amplitude (as defined in Equation 2), and $e$ is the planet's eccentricity, though in this work we set $e$=0 for all orbits.

\begin{equation}\
V_{\rm mod}(t) = K\left[\cos(f+\omega) + e \cos(\omega)\right]
\end{equation}\label{eqn:planetRV}

For the case of a single planet orbiting the star, the phase determination is straightforward. Things get slightly more complicated, however, when the star has two or more planetary companions. We adopt a simplified case where we disregard planet-planet interactions, so that the total instantaneous RV of a star being orbited by N planets is simply a linear combination of the RV semi-amplitude of each planet at the given time (Equation 4, where R is the noise factor).

\begin{equation}
V_{\rm star}(t) = \sum_{\rm i=1}^{N} V_{\rm mod,i}(t) + R(t)
\end{equation}\label{eqn:totalRV}

Combining these equations allows our scheduler to determine each planet's current phase curve location when deciding which star to observe. This, in turn, allows us to pair different planetary phases with higher or lower priority values depending on how valuable we think it would be to add a new data point at that location. In the remaining parts of this section we describe four different time-varying prioritization schemes that are then employed in our simulations. 

\subsection{Prioritization Schemes}

{\bf In and out of quadrature prioritizations}: The first two prioritization schemes are designed with the goal of trying to target or avoid the quadrature points of a planet's phase curve. The quadrature points (when the amplitude of the RV induced by the orbiting planet is at a maximum) reveal the semi-amplitude of a planet's RV curve, which can then be transformed into a measurement of its minimum mass value (\msini) or true mass value, depending on whether or not the planet's inclination is known.

\begin{figure}
\centering
\plotone{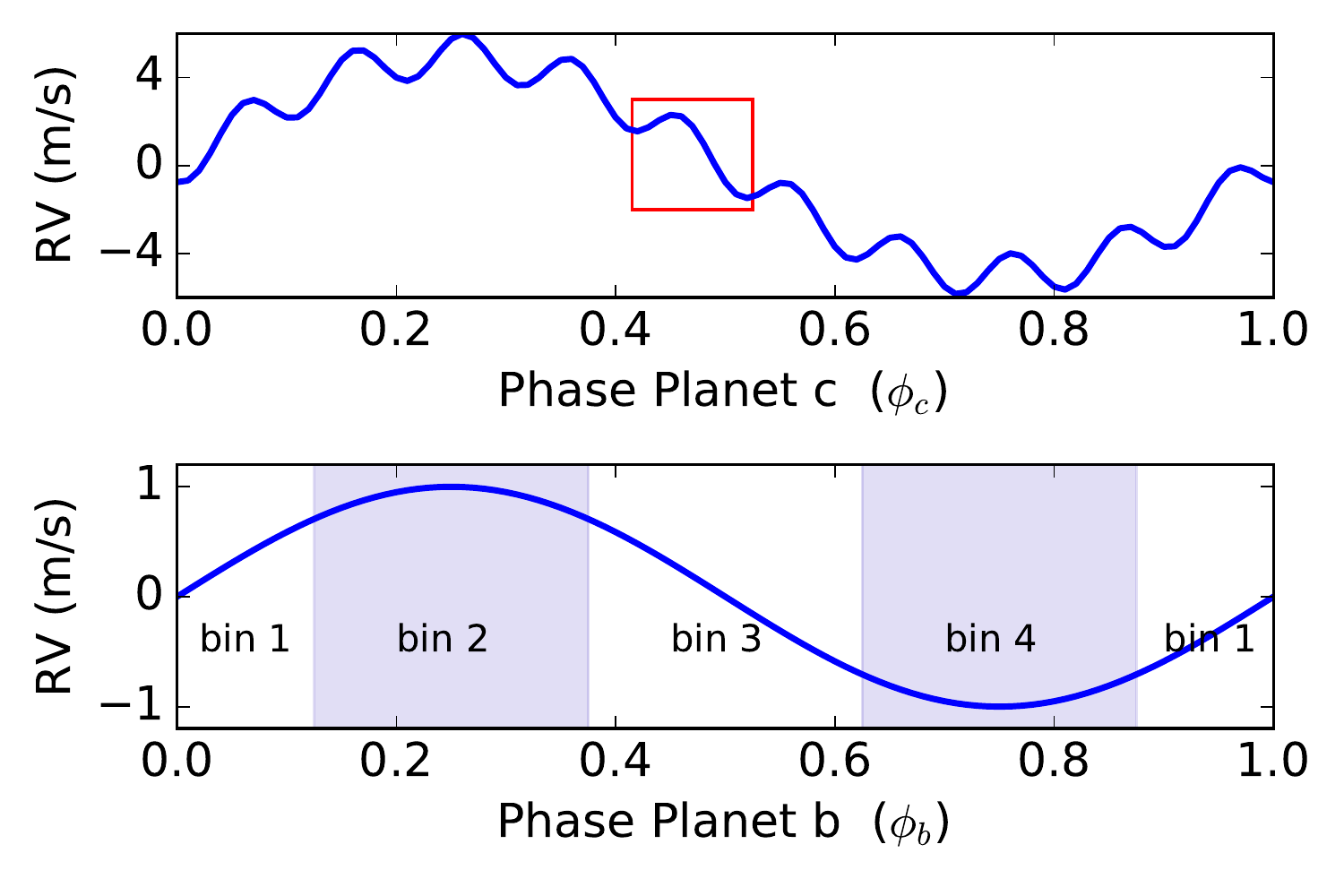}
\caption{\label{fig:inquadRVbins} {\it Top\ }: Schematic RV phase curve of two planets orbiting a star: a massive long period planet and a smaller, shorter period planet. The red box covers one orbit of the short period planet, which is depicted below after the longer period planet's signal has been subtracted. {\it Bottom\ }: In a multi-planet scenario, the in quadrature prioritization systems starts by folding on the shorter period planet's phase curve and defining the phase bins. The blue bins are centered on the quadrature points of the selected planet's orbit, while the white bins are centered on the zero-crossing points.}
\end{figure}

We define four, 90$^{\circ}$ bins across the folded phase curve of a planet, two of which are centered at the planet's quadrature points and two of which center on the zero-crossing points. For multi-planet systems, we select the shortest period planet and use its period and epoch of mid-transit to determine where in phase space the star falls at the time of a given observation (Figure \ref{fig:inquadRVbins}).

For our in quadrature prioritization scheme, stars currently in the light blue, quadrature sections (bins 1 and 3) of Figure \ref{fig:inquadRVbins} are assigned a priority of 10, while those in the zero-crossing bins (bins 2 and 4) are assigned a priority of 5. 

The out of quadrature scheme is exactly the opposite of the in quadrature approach, setting the priority to be 10 when the planet is in one of the zero-crossing bins and to 5 when the planet is in a quadrature bin.

{\bf Uniform prioritization}: This approach samples the phase curve of each star as uniformly as possible, in addition to sampling the catalog of targets as uniformly as possible. The scheduler computes the current RV phase for every available target when a new observation is to be performed. The priority of each star is based upon how far an observation taken at that moment would be from the nearest observation in the phase curve. The phase is normalized by the orbital period, which comes from the transit catalog, and then multiplied by 10, providing a number between 0 and 10. If there are no previous observations, the star gets the maximum possible priority. 

{\bf Random prioritization}: This approach, as the name suggests, employs a random sampling scheme. Every time a new target is required, we use a random number generator to select a star. The only requirement for a target to be selected is that it must remain above the elevation limits of observing long enough to complete the requested observation.

\subsection{Further considerations}

After the initial assignment of priorities, we assign a +1 in priority to any stars hosting planets with radius, R \textless 4\rearth\ in all schemes except the random one. This emphasizes our scientific interest in small planets, and ensures that the APF is helping to contribute to TESS' level one science requirement of measuring the masses of 50 small exoplanets.

Our priority assignment process also allows for the insertion of stars that have not been previously observed by missions like {\it TESS}, and thus do not have accurate transit based phase information readily available. As an example, RV standard stars, which we use to monitor the continued performance of the APF, can be inserted into the observing database with static priorities depending on how often we want them to be observed. Stars from legacy RV surveys can be added in with static priorities if their possible planetary systems are not yet well defined, or with the phase information of a suspected planet to make sure its viability is assessed in an efficient manner. 

One differentiating aspect of these prioritization schemes that is not represented here is that schemes with more phase coverage (uniform and random) should be better able to measure each planet's orbital eccentricity. Our adoption of the S15 all circular orbits approach means that we do not suffer from the measurable effects that eccentricity is known to have on translating semi-amplitudes to actual mass values \citep{Ford2008, Loredo2011}. Thus the expected improvement from having well constrained orbital eccentricities is not considered when comparing how the different schemes perform, though it is certainly something that should be incorporated in future works.

\section{Simulated APF observing} \label{sect:simulator}

The values calculated for each planet described in the sections above (true mass, estimated mass, and desired RV precision) are combined with the host star and planet parameters from the {\it TESS} simulation (Vmag, planetary radii, RA, Dec, and B-V color) and placed in a new observing database for the telescope. We add additional columns for the star names, the planetary period on which to base the phase bins of each star, and the initial phase of the selected planet at a specified JD zero point. This database is read in by the observing software at the beginning of each simulated night and guides the software's decision making process throughout the night as it chooses what star to observe at a given time.

\subsection{Simulating the observations}

In normal operations, the APF selects its next target by picking the star with the highest (static) priority that can be observed in a reasonable amount of time (i.e. that can achieve its desired precision level in less than one hour of total observing). We have written a simulator that replicates this. The details of the actual observing process, and the relations used to compute exposure times, are described in B15, so we will simply sketch out the core steps here. For a given star, we compute the total exposure time needed to achieve its desired precision. The inputs for the exposure time calculation include the star's magnitude and color, and also the atmospheric seeing and extinction (due to cloud cover), which are based on measurements from the previous observation. We then use these values to estimate the exposure time for the next observation. Because we want to ensure a certain quality of signal, the total exposure time is almost always an overestimate and we use the exposure meter to determine the actual end of an exposure.

The scheduler starts each simulated night at -9 degree twilight by selecting and observing a B star from of a curated list of bright stars with high rotational velocities. The resulting observation is used for data reduction purposes, and to measure the first set of seeing and transparency values for that night. After the B star has been observed, the scheduler begins selecting targets from the observational database, which contains all of our {\it TESS} science targets. The selection process continues on repeat until there are no additional stars that can be observed before the sun reaches 9 degree dawn at the end of the night. 

At the beginning of a run, the simulator selects a random distribution of seeing and extinction values modeled using the past three years of APF operations. The extinction is modeled as multiplicative factor on the exposure time, so it is inversely proportional to the photon arrival rate. Each night uses a different mean and variance from which random deviates are drawn. The mean and variance of both the seeing and the extinction are correlated as found in the historical data. Therefore, we have ``good'' nights, where the seeing and extinction are both drawn from small means with small variances, ``ok'' nights where either the seeing or extinction is small but the other is large, and ``bad'' nights where the seeing is large, and the extinction can be so bad that the telescope never observes a target. For each simulated observation, we draw one of the seeing and extinction pairs from the parent distributions to compute the actual photon arrival rate. These random deviates are used to compute the exposure time for the next observation. Once a star is selected, a new seeing and extinction pair are drawn and used to compute the photon arrival rate for the simulated observation. Included in this process is our empirical model for the seeing as a function of elevation. 

For each observation, we record the total number of photons simulated to land on the exposure meter and in the iodine region (5000 - 6200 \AA) of the Levy spectrograph, using the relations from B15. We compute the mean from those relations and then draw a random deviate using the scatter listed. The number of counts in the iodine region gives an estimate of the expected precision, from Equation 1 or Figure 5 of B15. We use that as a mean and use the measured scatter (also from Figure 5 of B15) to compute the actual precision for the simulated observation. For each measurement we add in quadrature the internal uncertainty, the expected jitter for the star (see \S \ref{sect:jitter}), and the baseline instrumental noise found in B15. This combination yields the total error estimate. 

At the end of each exposure, the simulator combines the true mass values for all planets in the system (whether or not they were originally detected by {\it TESS}) with the JD of the exposure's photon weighted mid-point, and the planets' initial phases at the JD zero point to calculate the star's total, instantaneous radial velocity value. We then add the noise factor for the observation and append the resulting values into the star's velocity file, which can be opened in the publicly available Systemic console \citep{Meschiari2009} or a python analysis package like Radvel \citep{Radvel2017} to fit the data and determine the planetary masses.

The scheduling of the nights is random. At present, the APF telescope is used for a number of programs by a variety of institutions, some of which require specific cadence or even specific nights. To simulate this, we assumed that 40\% of the nights each year are available for {\it TESS} follow up, excluding the annual winter shut down. Nights are assigned in on/off pairs, where the simulator observes for two nights in a row and then draws a number of nights off from a uniform distribution of 2/3/4 nights. 

\subsection{Fitting simulated RV data} \label{fittingRVdata}

We use an automated fitting process built on the publicly available Systemic console \citep{Meschiari2009} to estimate the semi-amplitudes and corresponding mass values for each of the transiting planets detected by {\it TESS}. The data are first binned, as is the case with real radial velocity data, though instead of using the common 2 hour binning scheme, we instead group velocities within a width of no more than 10\% of a planet's orbital period. The data are binned within 10\% only for observations taken on the same night. So while this could span a much longer (or shorter) period of time than the traditional two hour binning approach, it will never span multiple nights.

The list of binned velocity values is then loaded into Systemic and all values known from the {\it TESS} data (such as the period, and the mean anomaly at periastron) are entered into the system kernal and frozen so that the fitting algorithm can not alter them. We also freeze the orbital eccentricity and the dataset zero-point parameters to 0, to help ensure orbit stability. To provide the fitting algorithm with a sensible first guess, we use the $M{_{\rm est}}$ parameter described in \S \ref{sect:planetmasses} as a starting value for the planetary mass(es). Once all of the parameters are entered for each planet detected by {\it TESS}, we then perform a single minimization using the Simplex algorithm. This minimization provides the best fit semi-amplitude, K, and the best fit mass estimate in Earth masses, calculated using the host star's mass from our input database. To estimate the standard deviation of the best fitting $K$ values, referred to as $\sigma_{\rm K}$ throughout the paper, we apply Systemic's built in Monte Carlo Markov Chain error analysis module to the binned data using four chains and a convergence criteria of GR \textless 1.05, where GR is the Gelman-Rubin statistic \citep{GelmanRubin1992}.

\section{APF Simulator results} \label{sect:results}

The final data sets for each prioritization scheme span 36 months of simulated observing, during which we obtain $\sim$360 nights of telescope time. For each scheme we run 10 different 36 month-long simulator iterations (resulting in a total of 40 three year observing campaigns) in order to quantify the scatter in our results that is due to randomness in our simulator.

\subsection{Adhering to the correct phases}

\begin{figure}[h]
\centering
\includegraphics[width=.5\textwidth]{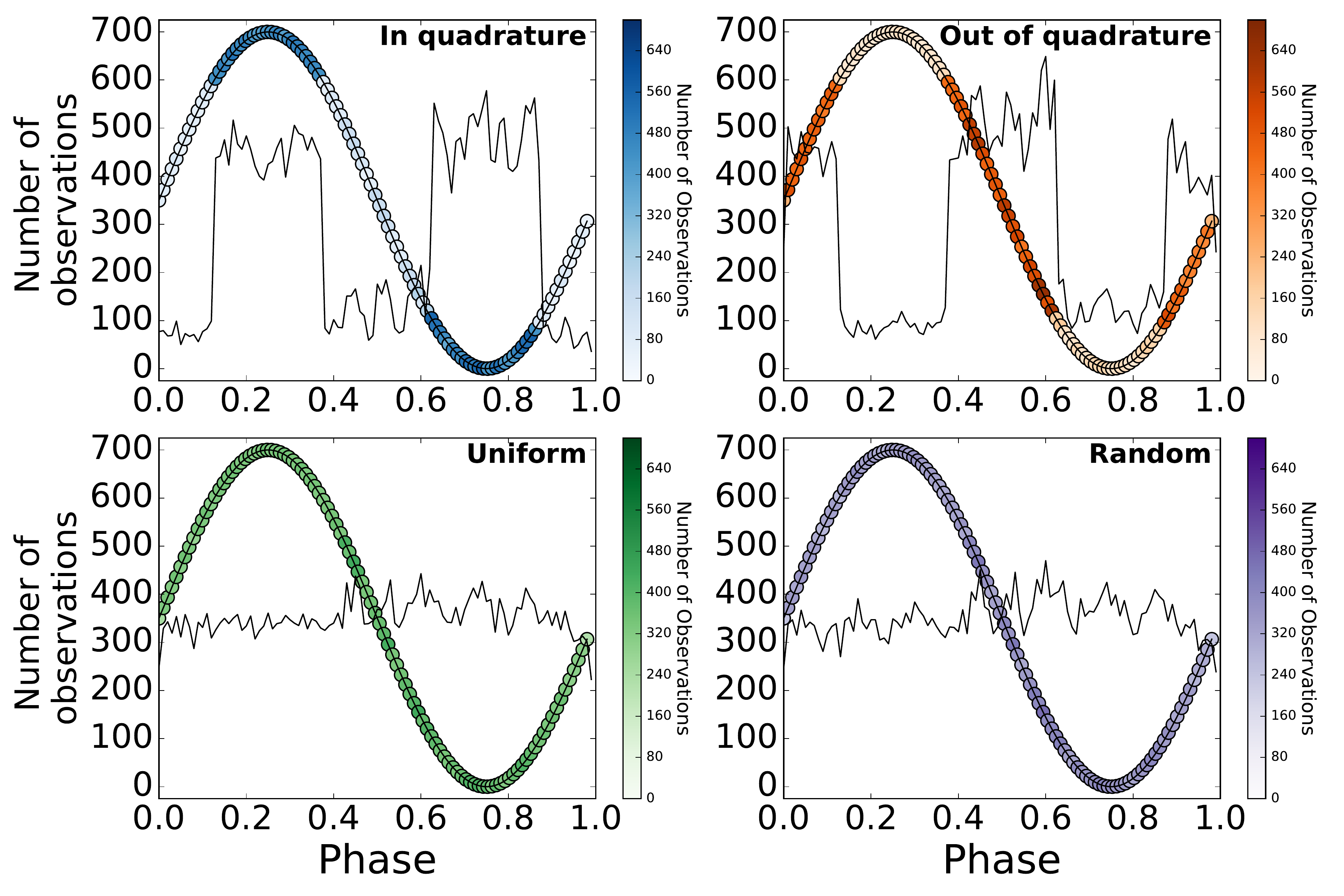}
\caption{\label{fig:phasemap} Phase coverage based on every observation taken across all 10 trials for each prioritization scheme. The clear increase of observation density on the in and out of quadrature schemes, along with the corresponding lack of over densities in the uniform and random schemes, shows that our scheduler is selecting targets as intended.}
\end{figure}

An important first check on these new observing strategies is to determine whether our four different prioritization schemes are actually sampling the planets' RV phase curves as intended. For each prioritization scheme we look at every observation across every target over all ten survey trials, and create a histogram of where in the planets' RV phase curves the data were taken (Figure \ref{fig:phasemap}). The results confirm that our prioritization schemes are acting as expected, with the in quadrature scheme showing a strong preference for observations happening near the quadrature peaks, the out of quadrature scheme showing a focus on the areas surrounding the planet's zero-crossing points, and the uniform and random schemes producing observations that are well spread throughout the phase curve.

\subsection{Number of planets detected by scheme}\label{sect:massmeasurements}

As described in Section \ref{sect:simulator}, each iteration run by the simulator uses the same input database, which contains information on the {\it TESS} planet candidates and their host stars, to guide the simulator's target selection process throughout the nights/years. Therefore any difference in the number of planets detected by each scheme is driven by the simulator's minute-by-minute prioritization decisions, under the constraints of random weather, not by differences in input the target list.

We randomly select one set of simulations as the data set used to create the Figures 7, 8, and 11 in this section, and from that simulation we incorporate only those stars that have N $\geq$ 10 observations. This value is used because lengthy experience in working with the \citet{Butler1996} I$_{\rm 2}$ reduction pipeline shows that the ``vanking'' procedure (which identifies and removes outlying 2$\AA$ wide chunks of the various stellar spectra before performing the Doppler RV analysis) produces more consistent results once the number of observations reaches at least 10. We also eliminate planets where the best fit masses have semi-amplitudes, K \textless \ 0.25 \ms. These are the result of a fitting artifact from Systemic, wherein it fits RV data sets that show no evidence of a Keplerian trend as having a planet with K $\sim$ 0 \ms\ and artificially small errors of $\sigma_{\rm K} \sim$ 0 \ms.  To date, no existing RV facility has detected a planet with K \textless \ 0.25 \ms.

The criteria that we use for assessing the significance of a given planet's mass measurement is the ratio of the measured semi-amplitude to the error in that semi-amplitude: K / $\sigma_{\rm K}$. When looking at the ensemble results, we find that systems where $K$/$\sigma_{\rm K} \ \textless \ 1$ show evidence of unrealistic mass measurements. We attribute this to a poor description of the data by the model. The best-fitting model does not take into account the periodic stellar noise contributions, nor any non-transiting planets in the system, both of which can distort or completely drown out the signature of the planet that we're actively trying to fit \citep{Rajpaul2017, Hall2018}. In general, for these systems, an approach that allows for more flexible fitting, such as fitting additional large peaks in the periodogram to help address stellar activity or additional planets, would likely help to clean up the fit and better constrain the mass of the transiting planet. Because of this, the results used in this section take into account only those planets for which $K$/$\sigma_{\rm K} \ \geq \ 1$, unless otherwise noted (see \S \ref{sect:MR_bias}). This cut is made to mimic real world analyses, in which any signal with $K/\sigma_{\rm K} \ \textless \ 1$ and without a large number of observations would likely be regarded with skepticism in the community.

\begin{figure}[h]
\centering
\includegraphics[width=.47 \textwidth]{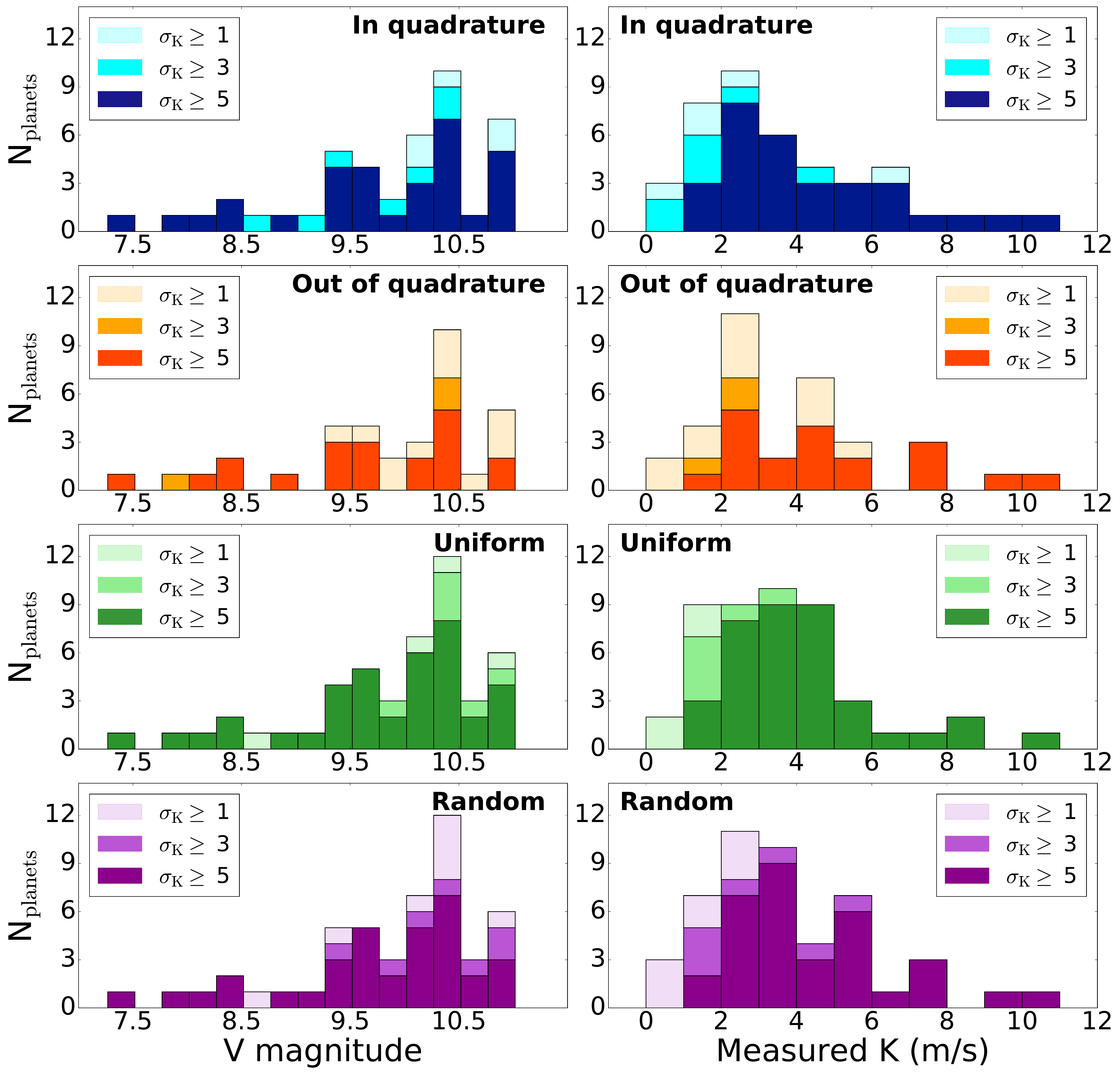}
\caption{\label{fig:SigmaHist} Results from one 36-month trial of simulated observing on the APF using each of the four time-varying prioritization schemes shown as a function of host star V magnitude (left column) and of measured planet semi-amplitude (right column). There is a lack of strong, distinguishing features between the different prioritization schemes.}
\end{figure}

Over the course of ten simulations, the four prioritization schemes produce the following results (Figure \ref{fig:SigmaHist}):

{\bf In quadrature:} The in quadrature simulations observed an average of 33.3 $\pm$ 1.1 stars for which it obtained at least 10 observations. Our fits to the resulting RV data sets produce 41.4 $\pm$ 1.0 planet mass measurements at the 1$\sigma_{\rm K}$ level, 35.4 $\pm$ 2.1 planet mass measurements at the 3$\sigma_{\rm K}$ level, and 28.3 $\pm$ 1.4 planet mass measurements at the 5$\sigma_{\rm K}$ level.

{\bf Out of quadrature:} The out of quadrature simulations observed an average of 33.1 $\pm$ 1.1 stars for which it obtained at least 10 observations. Our fits to the resulting RV data sets produce 35.6 $\pm$ 1.2 planet mass measurements at the 1$\sigma_{\rm K}$ level, 26.4 $\pm$ 1.9 planet mass measurements at the 3$\sigma_{\rm K}$ level, and 20.3 $\pm$ 0.9 planet mass measurements at the 5$\sigma_{\rm K}$ level.

{\bf Uniform:} The uniform simulations observed an average of 51.0 $\pm$ 0.0 stars for which it obtained at least 10 observations. Our fits to the resulting RV data sets produce 49 $\pm$ 1.3 planet mass measurements at the 1$\sigma_{\rm K}$ level, 44.9 $\pm$ 0.9 planet mass measurements at the 3$\sigma_{\rm K}$ level, and 36.9 $\pm$ 1.5 planet mass measurements at the 5$\sigma_{\rm K}$ level.

{\bf Random:} The random simulations observed an average of 44.6 $\pm$ 0.8 stars for which it obtained at least 10 observations. Our fits to the resulting RV data sets produce 47.3 $\pm$ 1.2 planet mass measurements at the 1$\sigma_{\rm K}$ level, 40.1 $\pm$ 2.6 planet mass measurements at the 3$\sigma_{\rm K}$ level, and 31.3 $\pm$ 1.3 planet mass measurements at the 5$\sigma_{\rm K}$ level.

When comparing the number of mass measurements made at the 1-, 3-, or 5$\sigma_{\rm K}$ level across all ten trials, we find statistically significant differences between the four prioritization schemes. We perform a two-sample t-test on the number of detected planets over all 10 trials of each scheme, with the null hypothesis that the mean number of detected planets is the same across prioritization schemes.  We find that we are able to reject this null hypothesis with p-values of 0.0005 or lower, except when comparing the 3$\sigma_{\rm K}$ uniform and random results, whose t-test produces a p-value of 0.01.

From these results, we conclude that the uniform prioritization scheme (which produces the highest number of planet detections at all significance thresholds) is the optimal observing strategy, followed by the random scheme, and then the in quadrature scheme. The out of quadrature scheme performs notably worse than the other three, which is unsurprising as its phase coverage provides little leverage in constraining the minimum and maximum points of a planet's RV phase curve. 

Our finding that the uniform and random schemes both perform better than in quadrature, even in the $e$=0 case studied here, shows that choosing one of these two eccentricity-probing schemes does not require a trade off in overall planet detection performance. Thus while the uniform approach already appears to be the superior RV observing approach for {\it TESS} follow up, the benefits of both the uniform and random approaches will likely be even more pronounced when constraining the masses of true planets on eccentric orbits.

\subsection{Measurement distributions by scheme}
As shown in Figure \ref{fig:SigmaHist}, the distribution of measured planet semi-amplitudes is similar across prioritization schemes. The distribution of host star V magnitudes also does not appreciably change. To test the significance of the small differences visible between schemes, we employed a two-sample KS test on a trial-by-trial basis. We find that the resulting two-tailed p-values never drops below 0.58 for any comparison in V magnitude space, nor below 0.14 for comparisons as a function of measured planet semi-amplitude. Therefore, the null hypothesis that they come from the same underlying distribution cannot be rejected, for any combination of trials and prioritization schemes. This is not surprising, as the measured planets in each iteration are indeed drawn from the exact same input target catalog, but it does emphasize the fact that the different prioritization schemes do not skew the detection distribution in a statistically significant way.

\subsection{Comparing prioritization scheme efficiency}

\begin{figure}[h]
\centering
\includegraphics[width=.5 \textwidth]{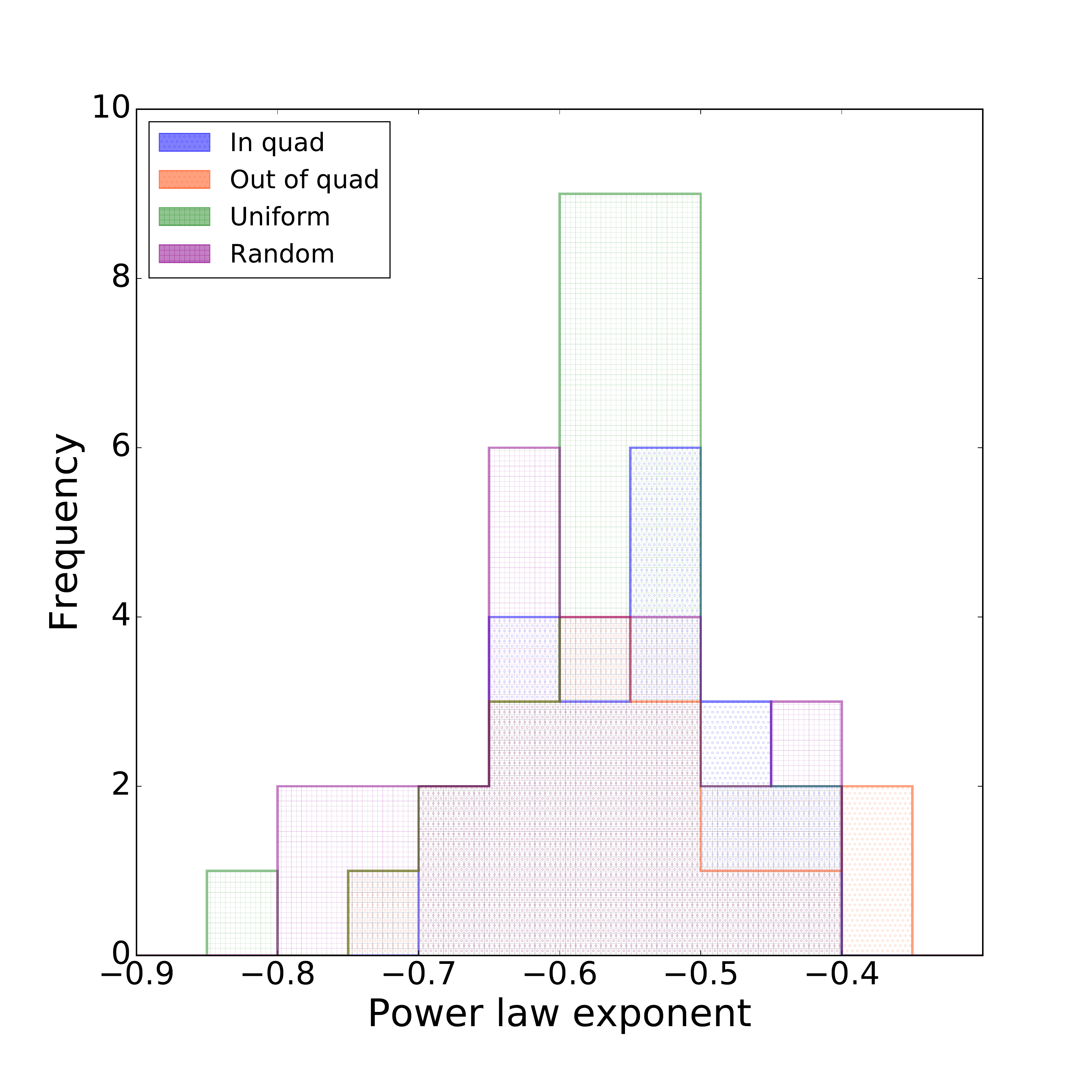}
\caption{\label{fig:powerlaw} Power law fits to the decrease in $\sigma_{\rm K}$ as a function of the number of RV data points (modeled as $\sigma_{\rm K}$ = 1/$N_{\rm obs}^{b}$ where $b$ is the shown power law index), sampled in three month intervals. The four schemes are not well differentiated, which implies that no single scheme reaches higher K/$\sigma_{\rm K}$ values faster than the others. In all four cases the mean power law fit is marginally steeper than the -0.5 value expected for the mean of a gaussian distribution, which is probably due to fitting a small number of RV points in the early data epochs.}
\end{figure}

In addition to comparing the number of 1-, 3-, and 5$\sigma_{\rm K}$ detections that each prioritization scheme produces, we also compare the rate at which they achieve these measurements. We select those stars that are observed in at least four different three-month segments throughout the three year observing simulation and that have K/$\sigma_{\rm K} \geq 1$. For these stars, we compare the error on the semi-amplitude measurement from Systemic against the number of RV data points in three-month intervals and fit a power-law of the form $\sigma_{\rm K}$ = 1/$N_{\rm obs}^{b}$ to the data. We find that the mean values of b are: -0.55$\pm$0.071, -0.56$\pm$0.092, -0.57$\pm$0.084, and -0.59$\pm$0.10 for the in quadrature, out of quadrature, uniform, and random schemes, respectively (Figure \ref{fig:powerlaw}). Thus all four prioritization schemes are marginally steeper than the 1/$\sqrt N_{\rm obs}$ behavior expected for the error on the mean of a gaussian distribution. This is likely due to the $\sigma_{\rm K}$ values being abnormally high in the early data, due to the small number of RV observations used to constrain the fit. As $N_{\rm obs}$ increases, the data conforms more to the 1/$\sqrt N$ behavior that we expect, and none of the prioritization schemes show evidence of improving their semi-amplitude measurement errors significantly faster than the others.

\subsection{Bias in individual small planet masses} \label{sec:individbias}

\begin{figure}[h]
\centering
\includegraphics[width=.5 \textwidth]{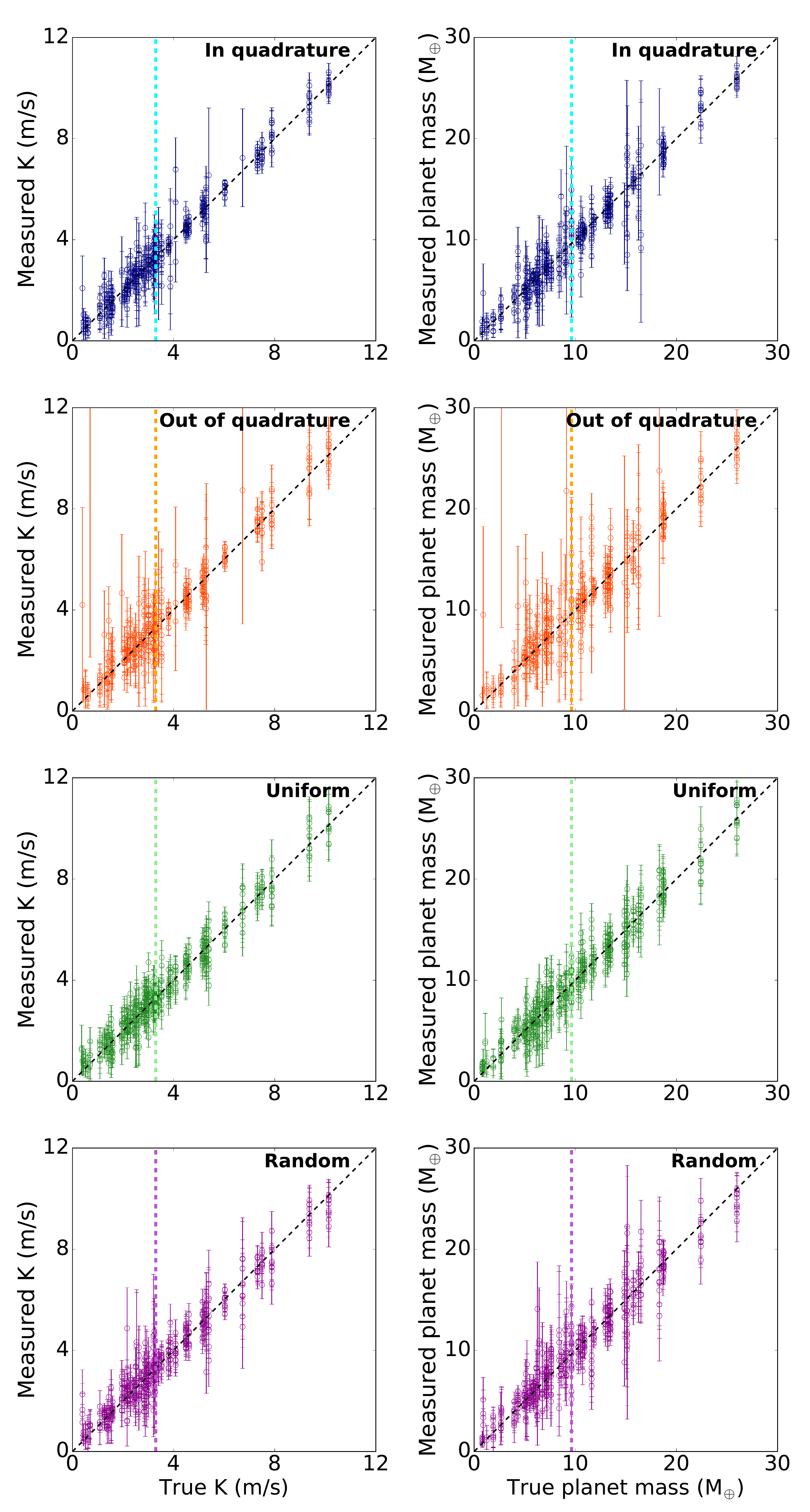}
\caption{\label{fig:MassCompare} Measured planet masses from our RV fits compared to the planets' true masses from the input catalog. The vertical dashed lines denote the median true semi-amplitude and mass values for all planets measured with at least 1 $\sigma_{\rm K}$ significance across all 10 trials, while the diagonal dashed line is a 1:1 relationship between the true and measured values.}
\end{figure}

Another interesting check that cannot be performed in real life is to see how our measured masses compare to the planets' true masses generated in \S \ref{sect:planetmasses}. Figure \ref{fig:MassCompare} shows the scatter of the measured masses compared to the planets' true masses around a dashed 1:1 line. The results of this comparison are summarized in Table 1, in the form of the mean, RMS and MAD values for the M$_{\rm meas}$ - M$_{\rm true}$ residuals. We split each prioritization scheme's results into two groups, based on whether the measured masses are higher or lower than the median true mass value of all planets which were measured at least at the 1 $\sigma_{\rm K}$ level across all 10 trials. Unsurprisingly, the mean of the residuals for planets above the median true mass is at least a factor of three smaller than for planets below the median true mass in all prioritization schemes, indicating that we obtain more accurate masses for planets that produce larger semi-amplitudes. Additionally, in all prioritization schemes we find that the smaller planets are more likely to have measured masses that are too large. This is likely due to a detection bias that comes from many of the small planets lying close to the K/$\sigma_{\rm K} \geq$ \ 1 detection threshold that we have set for inclusion in this analysis. For this group of planets, the uncertainty in their mass measurements means that we are more likely to include data from planets whose K values have been overestimated (moving them above our selection criteria) than data from planets whose K values have been under estimated (moving them below our selection criteria). This introduces a general bias of including small planets with measured masses that are higher than their true mass.

\begin{table}\label{table:mass_comp}
\caption{Fractional errors from measured planet masses vs true planet masses}
\begin{center}
\begin{tabular}{lcc}
\hline \hline
IQ residuals & M $<$ 9.6 M$_{\rm \oplus}$ & M $\geq$ 9.6 M$_{\rm \oplus}$ \\
\hline
mean & 0.061 & -0.010 \\
RMS & 0.403 & 0.107 \\
MAD & 0.218 & 0.057 \\

\\
\hline
OQ residuals & M $<$ 9.6 M$_{\rm \oplus}$ & M $\geq$ 9.6 M$_{\rm \oplus}$ \\
\hline
mean & 0.287 & 0.009 \\
RMS & 2.017 & 0.151 \\
MAD & 0.241 & 0.089 \\

\\
\hline
Uniform residuals & M $<$ 9.6 M$_{\rm \oplus}$ & M $\geq$ 9.6 M$_{\rm \oplus}$ \\
\hline
mean & 0.068 & 0.002 \\
RMS & 0.363 & 0.096\\
MAD & 0.209 & 0.076\\

\\
\hline
Random residuals & M $<$ 9.6 M$_{\rm \oplus}$ & M $\geq$ 9.6 M$_{\rm \oplus}$ \\
\hline
mean & 0.160 & 0.050 \\
RMS & 0.473 & 0.502 \\
MAD & 0.189 & 0.075 \\

\\
\hline \hline
\end{tabular}
\end{center}
\end{table}

\section{Mass-radius relations} \label{sect:MR_Results}

\begin{figure}[h]
\centering
\includegraphics[width=.5 \textwidth]{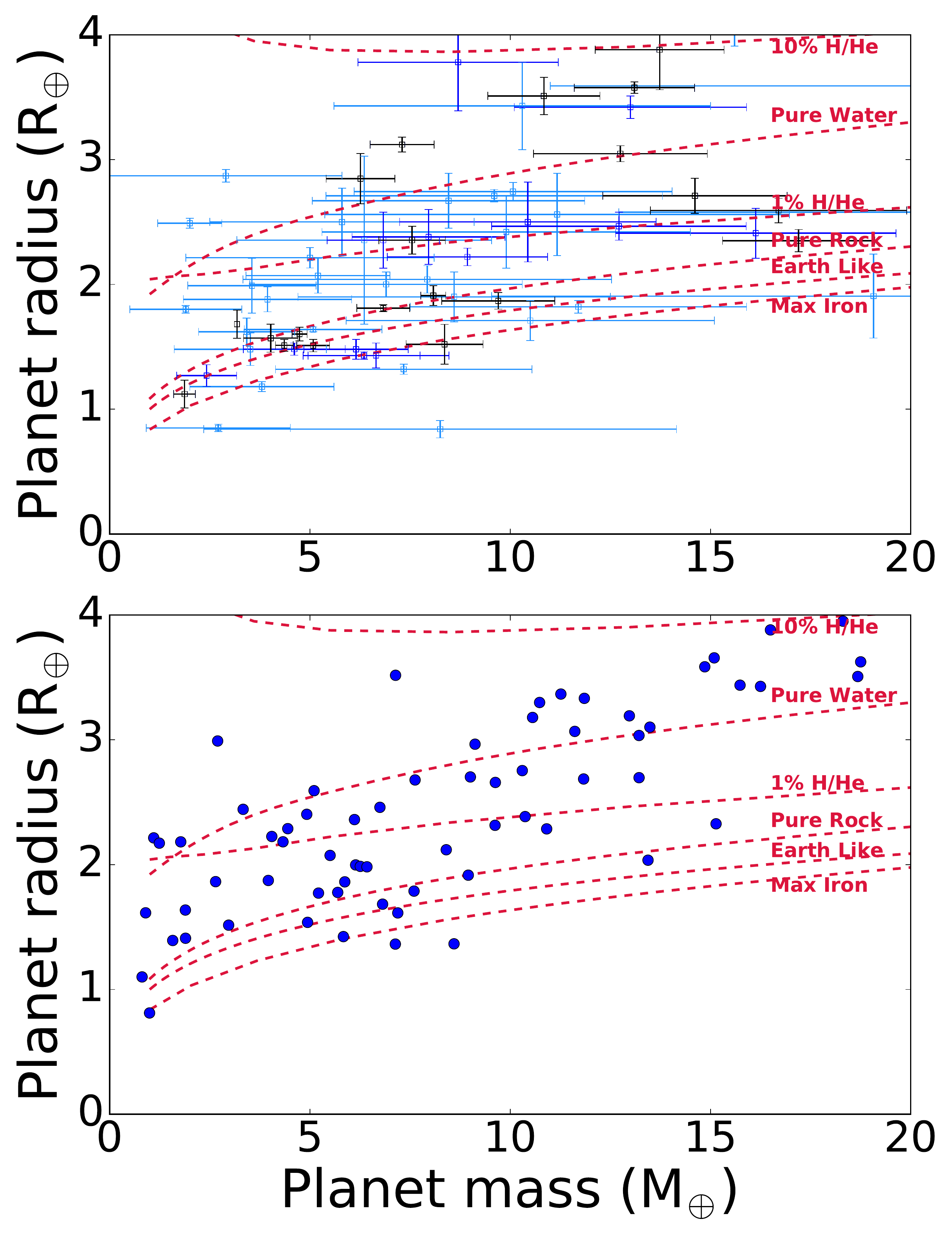}
\caption{\label{fig:CurrentMRD} {\it Top\ }: State of the exoplanet mass-radius diagram as of July 31, 2018, from the NASA Exoplanet Archive.  The composition curves come from \citet{Lopez2012} except for the ``max iron'' curve which is taken from \citet{Marcus2010}. The data points' colors represent their fractional mass uncertainties, with black points having uncertainties \textless 25\% of their measured masses, dark blue points having uncertainties from 25-50\% of their measured mass and light blue having uncertainties from 50-100\% of their measured masses. Our search criteria did not return any planets with fractional mass uncertainties \textgreater 100\%. {\it Bottom\ }: M-R diagram containing the planets we select from the simulated {\it TESS} results, showing the {\it TESS} radii and "true" masses derived using WRF16.}
\end{figure}

To establish context for the data points that {\it TESS} and the APF can add to the M-R diagram, we first create a current version using data from NASA's Exoplanet Archive as of September 26, 2018. After selecting those points that have both RV mass and radii measurements, along with error bars on both values, there are with 63 planets with radii, R \textless\ 4\rearth.  There are 27 planets whose total mass uncertainty is less than 25\% of their measured mass, another 25 whose total mass uncertainty is between 25-50\% of their total mass, and 11 planets with total mass uncertainties ranging from 50-100\% of their total mass,  all of which are plotted in Figure \ref{fig:CurrentMRD}. Our search criteria did not return any planets with fractional mass uncertainties \textgreater\ 100\%. The composition lines plotted on top of the M-R diagram are taken from \citet{Lopez2012}, except for the ``max iron'' line which represents the maximum iron fraction produced by collisional stripping according to \citet{Marcus2010}.

\subsection{Additions to the mass-radius diagram}\label{sect:MR_additions}

\begin{figure*}
\centering
\includegraphics[width=1. \textwidth]{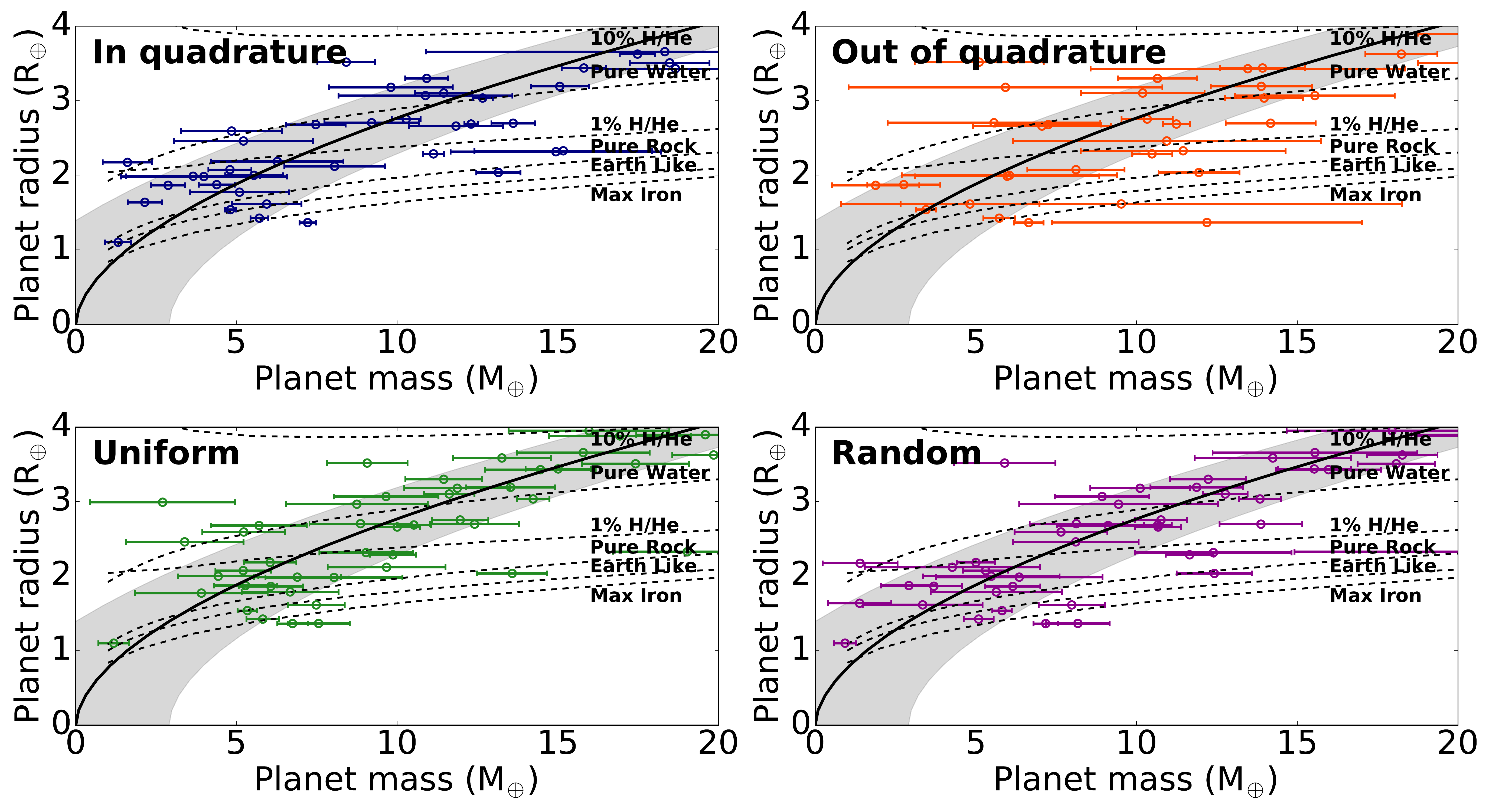}
\caption{\label{fig:APF_MRDs} Additions to the MR diagram from three-year surveys of TESS transiting planets with the APF, for each of our four new prioritization schemes. These plots show only those planets with mass measurements that have K/$\sigma \geq$ 1 from a single instance of simulated observing using each scheme, and include 43, 35, 48, and 49 new points for the in quadrature, out of quadrature, uniform, and random schemes, respectively. The black solid line and surrounding grey envelope depict our input M-R relation from WRF16, representing the mean relation and the standard deviation of the astrophysical scatter at a given radius.  The dashed black lines are the same composition curves as displayed in Figure \ref{fig:CurrentMRD}.}
\end{figure*}

Mass radius diagrams showing the K/$\sigma_{\rm K} \geq$ 1 additions from a single instance of each of the four different priority schemes are shown in Figure \ref{fig:APF_MRDs}. These plots contain the same data set displayed in Figure \ref{fig:SigmaHist}, and include 43, 35, 48, and 49 new points for the in quadrature, out of quadrature, uniform, and random schemes, respectively. 

\subsection{Underlying M-R relationships}\label{sect:MR_bias}

While obtaining more planet mass measurements is an important step in characterizing the exoplanets that we have discovered thus far, we would ultimately like to use this population to constrain theories of planet formation and evolution. One important factor for such work is the observed mass-radius relation, which is an empirical description of the exoplanet composition distribution. Unfortunately, the process of choosing which transiting planets to follow up with RV facilities, when to observe them, and when to publish the measured planet mass introduces biases into the planet sample. The scientific conclusions based on these samples, such as the typical rock/iron fraction of small planets \citep{Dressing2015} or the period dependence of exoplanet hydrogen envelope mass fractions \citep{Gettel2016, Sinukoff2017}, may then also be biased.   

Comparing observed vs. simulated planet populations can help illuminate the source of these biases and potentially help to mitigate them.
To do this, we create two versions of each simulated set, which consist of planet masses measured at different significance levels.  One includes only those planets with measurements at the K/$\sigma_{\rm K} \geq$ 3 level; this is meant to represent the planets that we would feel most secure about basing such analyses on in the real world.  The second includes all planets regardless of their K/$\sigma_{\rm K}$ value.

\begin{figure}
\centering
\includegraphics[width=.5 \textwidth]{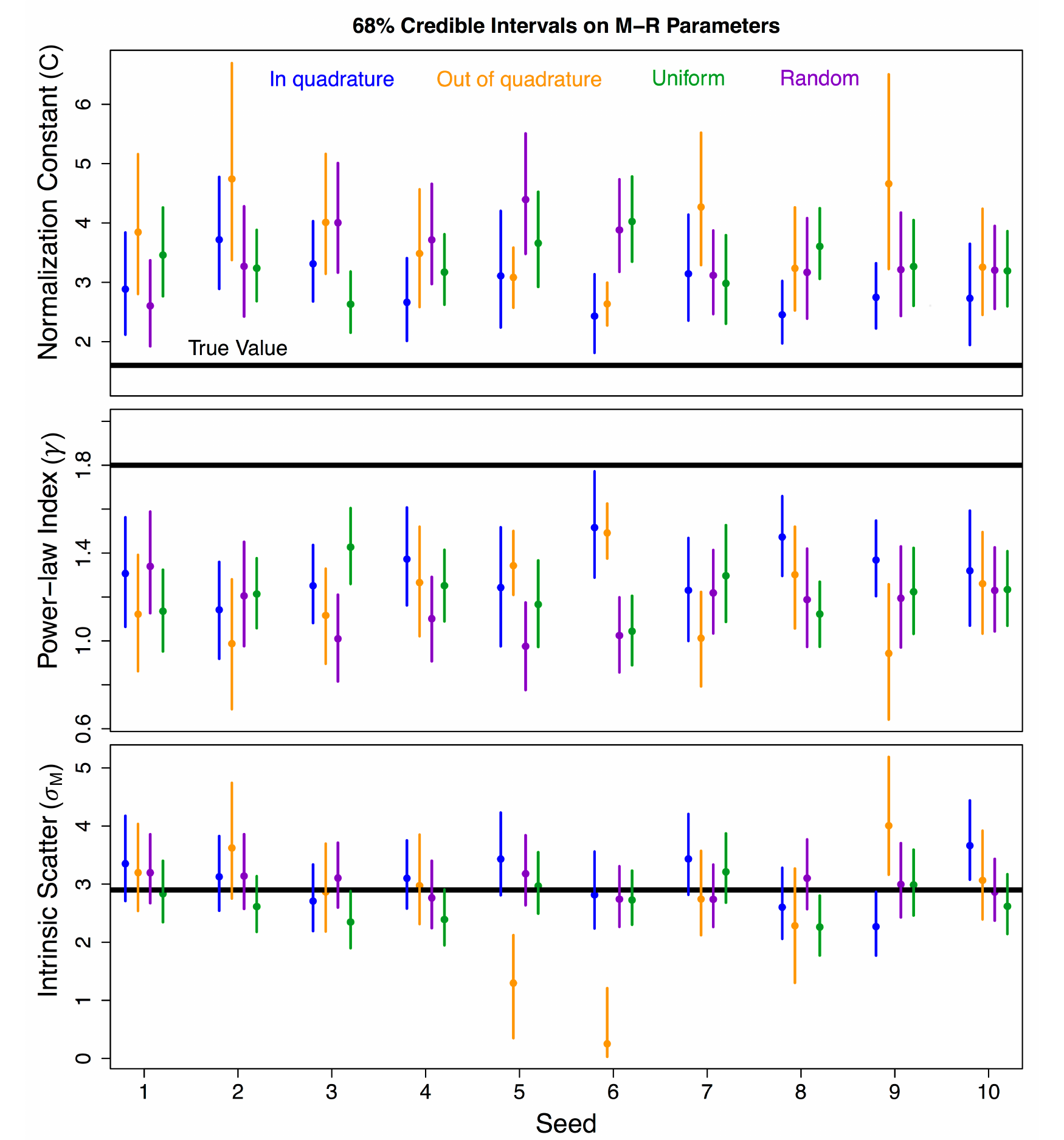}
\caption{\label{fig:MR_results1} The M-R relation parameters (see Equation 1) that were fit to each of the four prioritization schemes across all ten iterations of the simulator, using any planet masses where K/$\sigma_{\rm K} \geq 3$. The ``true" parameter values used to generate the initial data set of planet masses are denoted by the horizontal black lines. The normalization constant is consistently measured to be too high while the power-law index is consistently measured to be too low, meaning that the resulting M-R relation is too flat and too high compared to the truth. Therefore, the datasets produced by these simulations are biased toward higher-mass planets at small radii, independent of what prioritization scheme was used.}
\end{figure}

An efficient way to compare the input, true M-R relation to the observed one is to consider the results in parameter space. We separately analyze each simulated data set with the same hierarchical Bayesian method used to infer the original mass-radius relation of WRF16, which generated the true planet masses used in our RV analysis. As described in \S \ref{sect:planetmasses}, the input M-R relation takes the form shown in Equation 1, with parameters C = 1.6, $\gamma$ = 1.8, and $\sigma_{\rm M}$ = 2.9. We fit each simulated dataset using this functional form, allowing C, $\gamma$, and $\sigma_{\rm M}$ to vary. The resulting parameters for each iteration of each phase prioritization scheme are shown in Figure \ref{fig:MR_results1}.

As seen in the top panel of Figure \ref{fig:MR_results1}, the ``more trustworthy'' dataset with the K/$\sigma_{\rm K} \geq 3$ cut yields biased M-R relation parameters.  Specifically, the normalization constant C (defined to be the average mass of a planet at 1R$_\oplus$) is shifted above the input, true value across all 10 simulation iterations {\it and} all four observing schemes. This means that the inferred M-R relation falls above the ``true" relation from which the planet masses were generated, i.e. the planet population is inferred to be more massive than it actually is. 

The upwards bias in the normalization constant C is matched by a downward bias in the power-law index $\gamma$ due to the correlation between the two parameters (see Figure 2 of WRF16).  This arises because the larger planets' masses are well-measured: 
when the normalization constant is constrained to be too high, the power-law slope becomes flatter to compensate, in order to fit the less biased masses of the larger, well-measured planets.

Finally, the fits of the four prioritization schemes in the intrinsic scatter parameter (bottom panel of Figure 12) are on average consistent with the truth, with some of the out-of-quadrature iterations failing at the $\sim 1.5 \sigma_{\rm K}$ level. Therefore, the target prioritization schemes simulated here sample the full diversity of planet compositions, and the result is broadly independent of phase prioritization scheme.

\begin{figure}
\centering
\includegraphics[width=.5 \textwidth]{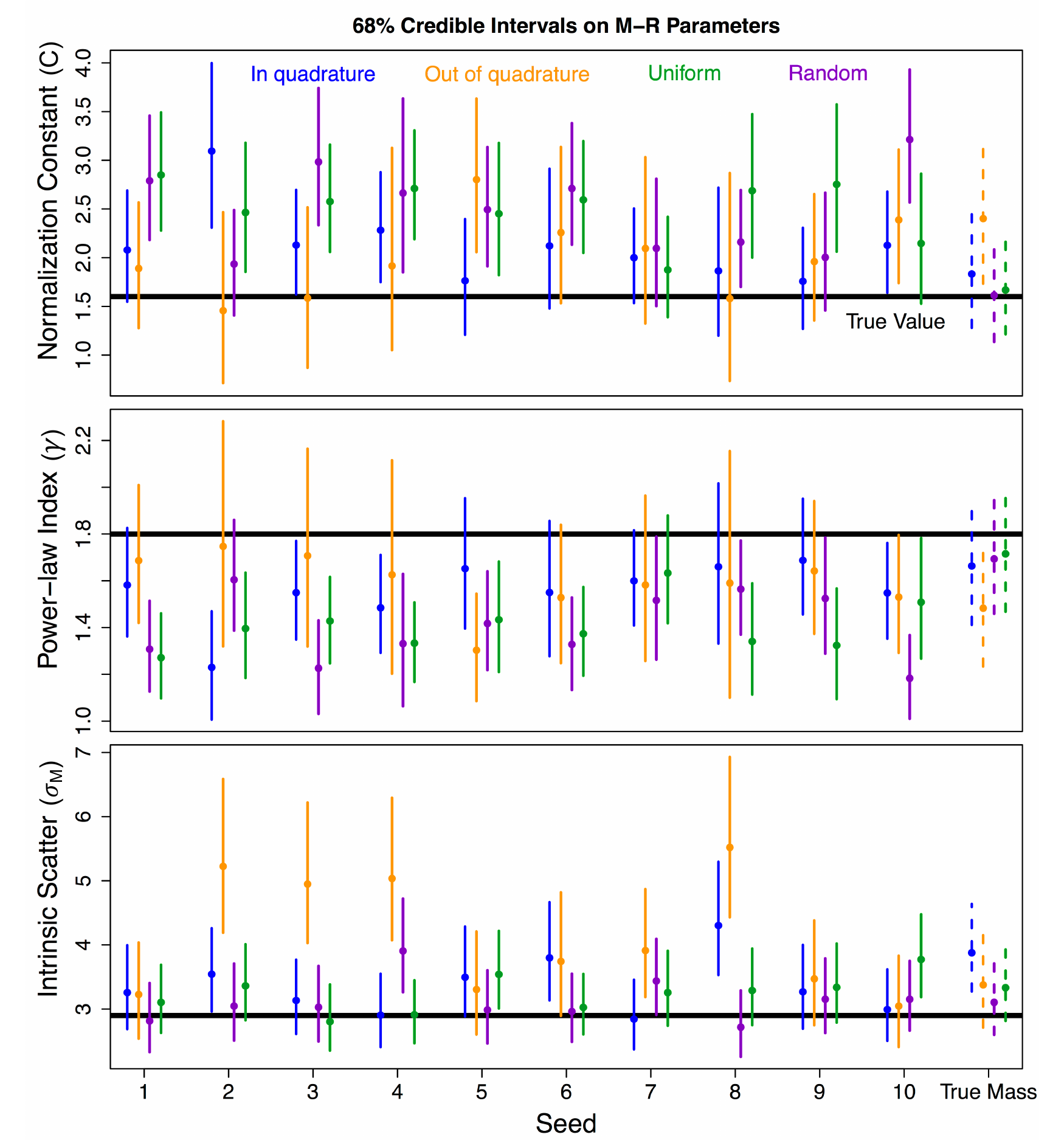}
\caption{\label{fig:MR_results2} The M-R relation parameters fit to each of the four prioritization schemes across all ten iterations of the simulator, but now without any K/$\sigma_{\rm K}$ cut.  The power-law index and normalization constant are much closer to the true values, thus confirming that the significance cut produced the majority of the bias in the M-R relation.  The small remaining bias is not due to the M-R relation fitting procedure: the dashed lines, which mostly span the truth, were fit to the \emph{true} masses of the planets that are observed in at least one instance of the simulator, for each prioritization scheme.  Therefore, there is still a bias between the observed population and the simulated population.  The residual bias could be due to a mismatch between our generated RV noise model and our analysis noise model.}
\end{figure}

What causes this bias?  Given the individual planet mass bias noted in \S \ref{sec:individbias} and Table \ref{table:mass_comp}, we suspected that the M-R relation is too high and too flat because only the more massive planets at a given size are included in the dataset used to constrain the M-R relation.  To test this, we performed the same analysis with the second version of our simulated dataset, which consists of all of the observed masses in each iteration, regardless of their K/$\sigma_{\rm K}$ value (Figure \ref{fig:MR_results2}). Extracting the observed M-R relation from all planets, even those that effectively only have mass upper limits, brings the power-law index and normalization constant closer to the true values.  Thus, the significance cut does indeed produce most of the bias in the M-R relation; this is consistent with the findings of \citet{Montet2018}.  While this bias affects all planets $< 4$R$_\oplus$ in our survey, its effect is strongest on the Earth-sized planets - those that we often focus on when considering questions about habitability and the occurrence rate of Earth-like bodies.  Indeed, for 1R$_\oplus$ planets, we generally infer a mass of $2.5 - 4 M_{\oplus}$. 

Even the full set of measured planet masses, however, does not bring us all the way to the true values. This residual bias is not due to our hierarchical modeling approach: when we construct a third set of data consisting of the \emph{true} masses of planets observed in any simulation trial, for each prioritization scheme, we find that the normalization constant and power-law index are determined correctly by the in quadrature, uniform, and random schemes (dotted lines, Figure \ref{fig:MR_results2}).  The continued bias in the out-of-quadrature scheme is likely due to the smaller data set produced by that scheme; smaller data sets are more sensitive to the stochasticity of which specific planets are included in any given realization.  The out-of-quadrature scheme is also the most poorly performing in our simulations.

If it's not the fitting scheme that produces this small, residual bias, then what does?  One possibility is that, to fit the 1500+ RV data sets in this work we adopted a simplified model for the RV jitter that did not account for the periodic stellar oscillations and rotation RV signals that we simulated in \S \ref{sect:jitter}. This mismatch between our generated model and our analysis model has likely produced an additional small bias in the observed planet masses compared to the truth.  A hint that this is the case lies in a comparison between the in-quadrature parameter constraints and the random or uniform constraints: the in-quadrature constraints are slightly closer to the truth than the others.  This could mean that the larger number of points close to the maximum planetary RV signal for this scheme better (but not completely) averages out the stellar noise in a region of phase space that is relatively important for constraining an accurate RV semi-amplitude.  
In any case, this effect is small given the randomness across trials, and we reserve investigation of this effect for future work.  Nevertheless, it highlights the importance of choosing an appropriate noise model for the RV signals induced by stellar activity.

Including all mass measurements, regardless of their K/$\sigma_{\rm K}$ value, in an M-R extraction analysis also raises concerns about the validity of low significance mass measurements like the small planet, K/$\sigma_{\rm K}$ \textless\ 1 data points that were found to have unrealistic masses in \S \ref{sect:massmeasurements}. Fortunately, most of these unrealistic masses have K \textless \ 0.25 \ms, so we are able to identify and remove these problem cases from the analysis here.  That said, the masses of planets which induce small semi-amplitude signals on their host stars should be treated with extra care to ensure that their measured masses are not influenced by other signals in the data, especially as our M-R extraction tests show the importance of including all planets in population analyses. Teams preparing M-R analyses should thus ensure that they have a thorough understanding of the fitting process used to extract masses from the RV data, and check that such planets undergo thorough vetting processes that look into potential physical (e.g. stellar activity and other Keplerian signals in the data) and software (e.g. the orbital parameter basis set used for fitting and the minimization algorithm used in the fit) mass bias sources.

While the datasets presented here were shaped by an observing simulator specific to the APF, many of the decisions incorporated into this simulator follow common practices in the field, and so these results should be broadly applicable to other RV groups. Our main takeaway from this exercise is that observing groups who aim to analyze the empirical mass-radius distribution produced by a given RV follow up effort need to include mass measurements of all planets in their sample, no matter the ``sigma'' value. Performing population analyses on \textgreater\ $X$-sigma masses will consistently produce biased results in the extracted M-R relation. Indeed, when examining previously published M-R relations which used all available RV masses as of 2015 \citep[e.g.][]{Wolfgang2016, Weiss2014}, we see predictions of 3\mearth\ planets at 1\rearth\ radii, which is likely evidence of this bias in action. In any case, we recommend that teams who, at any point, plan to look for trends in their population of measured masses and radii perform a similar investigation into the biases produced by their target selection and night-by-night observing protocols.

\section{Conclusion} \label{sect:conclusion}

The results presented here are the first assessment of the contributions that the Automated Planet Finder telescope could have when participating in RV follow up efforts for {\it TESS}. After 36 months of simulated observations using 40\% of the telescope's nights, during which we attained roughly 360 nights of data, our survey produced $\sim$\ 30 5$\sigma_{\rm K}$ mass estimates of planets originally detected by the {\it TESS} mission. We tested how the choice of phase prioritization (in quadrature, out of quadrature, uniform, or random) affects the number of {\it TESS} planet masses we can measure at 1-, 3-, and 5$\sigma_{\rm K}$ confidence and see that the uniform approach performs the best, consistently producing more mass measurements than the other schemes all levels. The random scheme performs second best, followed by in quadrature, and lastly out of quadrature. Additionally, no individual scheme stands out as being more efficient at reaching a given $\sigma_{\rm K}$ level than the others. 

These results suggest that the benefits of the uniform and random schemes - which have the phase coverage necessary to determine a planet's orbital eccentricity - could be even more pronounced when studying systems containing planets on non-circular orbits. Using one of the schemes that fills out the entire RV phase curve would allow for determining the planet's eccentricity without requiring additional observing time to reach the same K/$\sigma_{\rm K}$ significance level. But it is also likely that the performance of the various schemes {\it will} change for planets on eccentric orbits, so this should be studied in more depth.

We also compared our measured RV masses against the simulated `true' planet masses. Below roughly 10\mearth we see a systematic bias of measured masses that are higher than the true planet mass.  This bias comes as a consequence of only reporting masses for planets measured with high statistical significance. We tested how this bias affects the mass-radius relation inferred from our TESS + APF data set and find that it produces a significant bias in the M-R relation parameters: the recovered M-R relation is flatter, and pushed higher than it should be. When examining previously published M-R relations which used all available RV masses at their time of publication \citep[e.g.][]{Wolfgang2016, Weiss2014}, we see predictions of 3\mearth\ planets at 1\rearth\ radii, which is possibly evidence of this bias in action. 

Determining the likely composition of a planet hinges critically on measuring a precise and accurate mass. If the RV field's reporting approach has biased our understanding of the small planet mass-radius relation, then this affect could ripple forward in at least two ways. First, our understanding of the spread of small planet compositions in the galaxy will be skewed towards high density planets, and will not accurately reflect the true distribution of planet types.  Second, people will use the biased mass-radius relations to predict the masses of newly discovered transiting planets. These planets, with erroneously high predicted masses, are then likely to be targeted with RV follow up efforts that do not reach the precision level necessary to characterize the true (smaller) semi-amplitude of the planet. This could result in numerous observations that are not able to help constrain the planet's true mass, thereby wasting telescope time on the already heavily over-subscribed set of PRV instruments currently available for such work. If the community began reporting upper limits along with statistically significant detections on all the planet masses we attempt to measure then this bias and its associated affects could be mitigated.

Another takeaway from this work is our ability to test and evolve our observation strategy for {\it TESS} (and later {\it PLATO}) follow up. Having now developed this second generation dynamic scheduler and the software necessary to implement long-term survey simulations, we have the ability to investigate a variety of observing strategies in preparation for our RV follow up campaigns. Going forward, we need to test if our current approach, which maximizes the observability of a star, has any biases. We also need to determine the effects of notably different survey approaches (e.g. focusing on one small group of stars at a time and observing them multiple times per night to beat down the RV noise levels as in \citep{LopezMorales2016}) on the mass estimates we are able to generate. 

Our simulator can help to improve general RV survey planning and data assessment. In the current approach to RV science, decisions on target selection, data acquisition method, and science results on the detected planet demographics are tied tightly together and strongly influence on one another. By implementing a consistent and well defined algorithmic process for each of these steps and restricting that feedback loop, we can look into whether a survey that is unbiased by humans can gain a better understanding of nature's underlying distributions and the physics that shapes these planets. In particular, experiments that examine how changing the prioritization scheme or target selection process part way through the survey can affect the M-R relation inference could shed much needed light on the best ways to move forward with RV follow up efforts across many telescopes, not just the APF. Indeed, if other facilities perform similar assessments and map out their performance across the M-R diagram, noting what types of stellar hosts and planet types they are best suited to observing, it could provide a valuable step towards a coherent and well organized approach to obtaining masses for the planets detected by {\it TESS}.

In short, it is immediately clear that the APF can provide a valuable set of mass measurements for planets with masses M $<$ 4\mearth, where the underlying M-R relation is not yet well understood. Indeed, our measurement of $\sim$ 30 planets at the 5-$\sigma_{\rm K}$ level with the better observing schemes is a significant step towards {\it TESS}'s level one science requirement of measuring the masses of 50 small exoplanets.

{\it Facilities:} \facility{APF (Levy Spectrometer)}, \facility{Exoplanet Archive}

\acknowledgements{
JB acknowledges support from the MIT Kavli Institute as a Torres postdoctoral fellow.  AW’s financial support during this investigation was provided by the National Science Foundation under Award No. 1501440, with facilities and some logistical support provided by the Center for Exoplanets and Habitable Worlds.  The Center for Exoplanets and Habitable Worlds is supported by the Pennsylvania State University, the Eberly College of Science, and the Pennsylvania Space Grant Consortium.  Portions of this research were conducted with Advanced Cyber Infrastructure computational resources provided by The Institute for CyberScience at The Pennsylvania State University (https://ics.psu.edu).

The work herein was based on observations obtained using the Automated Planet Finder (APF) telescope and its Levy Spectrometer at Lick Observatory.

This research has made use of the NASA Exoplanet Archive, which is operated by the California Institute of Technology, under contract with the National Aeronautics and Space Administration under the Exoplanet Exploration Program.}

\bibliographystyle{apj}

\end{document}